\documentclass[12pt]{article}
\usepackage{epsfig,amsmath}
\makeatletter
\renewcommand{\section}{\setcounter{equation}{0}\@startsection
  {section}%
  {1}%
  {0pt}%
  {-1\baselineskip}%
  {0.4\baselineskip}%
  {\large \bfseries}}%
\renewcommand{\subsection}{\@startsection
  {subsection}%
  {2}%
  {0pt}%
  {-0.75\baselineskip}%
  {0.2\baselineskip}%
  {\bfseries}}%
\renewcommand{\subsubsection}{\@startsection
  {subsubsection}%
  {3}%
  {0pt}%
  {-0.5\baselineskip}%
  {0.1\baselineskip}%
  {\sc}}%
\makeatother

  {\end{list}}%

\def\a{\alpha} 
\def\b{\beta}
\def\d{\delta}

\def\gm{\Gamma}
\def\e{\eta} 
\def\la{\lambda}        
\def\La{\Lambda}

\def\m{\mu}
\def\n{\nu}
\def\r{\rho}
\def\s{\sigma}

\def\th{\theta}

\def\ee{\varepsilon}
\def\om{\omega}

\def\tp{\tilde{p}}
\def\tk{\tilde{k}}

\def\vs{(\tilde{p}_1+\tilde{p}_2)^2}
\def\vt{(\tilde{p}_2+\tilde{p}_3)^2}
\def\vu{(\tilde{p}_1+\tilde{p}_3)^2}
\def\sumlogs{\sum_{i=1}^4\, \ln \tp_i^2}

\def\cosppp{\cos\!\left(\frac{p_1\wedge p_2 + p_1\wedge p_3 
                              + p_2\wedge p_3}{2}\right)}
\def\cosppm{\cos\!\left(\frac{p_1\wedge p_2 + p_1\wedge p_3 
                              - p_2\wedge p_3}{2}\right)}
\def\cospmm{\cos\!\left(\frac{p_1\wedge p_2 - p_1\wedge p_3 
                              - p_2\wedge p_3}{2}\right)}

\def\etaonetwothreefour{\eta_{\m_1\m_2}\eta_{\m_3\m_4}}
\def\etaonethreetwofour{\eta_{\m_1\m_3}\eta_{\m_2\m_4}}
\def\etaonefourtwothree{\eta_{\m_1\m_4}\eta_{\m_2\m_3}}

\def\idpd{\int\! \frac{d^D\!p}{(2\pi)^D} \,\,}
\def\idq{\int\! \frac{d^4\!q}{(2\pi)^4} \,\,}
\def\idqd{\int\! \frac{d^D\!q}{(2\pi)^D} \,\,}
\def\idx{\int\! d^4\!x \,}
\def\idxd{\int\! d^D\!x \,}

\def\ds{\displaystyle}
\def\to{\rightarrow}

\def\RR{{\rm I\!\!\, R}}

\begin{document}
\begin{titlepage}
\rightline{UCM-FT/00-13-01}
\rightline{hep-th/0007131}

\vskip 1.5 true cm
\begin{center}
{\Large \bf Paramagnetic dominance, the sign of the beta function\\[9pt]
and UV/IR mixing in non-commutative $\boldsymbol{U(1)}$}\\ 
\vskip 1.2 true cm 
{\rm C.P. Mart\'{\i}n}\footnote{E-mail: carmelo@elbereth.fis.ucm.es}
{\rm and} 
{\rm F. Ruiz Ruiz}\footnote{E-mail: t63@aeneas.fis.ucm.es}\\ 
\vskip 0.3 true cm
{\it Departamento de F\'{\i}sica Te\'orica I}\\
{\it Facultad de Ciencias F\'{\i}sicas}\\ 
{\it Universidad Complutense de Madrid}\\
{\it 28040 Madrid, Spain}\\
\vskip 1.2 true cm

{\leftskip=45pt \rightskip=45pt 
\noindent
$U(1)$ gauge theory on non-commutative Minkowski space-time in the
Feynman-'t Hooft background gauge is studied. In particular, UV
divergences and non-commutative IR divergent contributions to the two,
three and four-point functions are explicitly computed at one loop. We
show that the negative sign of the beta function results from
paramagnetism --producing UV charge anti-screening-- prevailing over
diamagnetism --giving rise toUV charge screening.  This dominance in
the field theory setting corresponds to tachyon magnification
dominance in the string theory framework.  Our calculations provide an
explicit realization of UV/IR mixing and lead to an IR renormalization
of the coupling constant, where now paramagnetic contributions produce
screening and diamagnetic contributions anti-screening.  \par }
\end{center}

\vfil

\end{titlepage}
\setcounter{page}{2}

%----------------------------------------------------- Paper

\section{Introduction}

It is a long established fact that the negative sign of the one-loop
beta function of QCD can be beautifully explained as the result of a
competition between the paramagnetic and diamagnetic contributions to
the bare coupling constant obtained from integrating out very
high-energy degrees of freedom. Indeed, the paramagnetic contribution,
which anti-screens the charge, prevails over the diamagnetic
contribution, which screens the charge \cite{antiscreen1}
\cite{antiscreen2}. This dominance can be explicitly computed by
evaluating the high-energy quantum fluctuations around a classical
field configuration \cite{Polyakov}, which in turn can be done in a
very elegant way using the background field method \cite{Abott}.

The vacuum-to-vacuum amplitude in the presence of an $SU(N)$
background field $B_{\m}$ is given by the partition function $Z[B]$,
which up to one loop, reads in the Feynman-`t Hooft background field
gauge 
\begin{equation} 
      Z[B]= e^{iS_{\rm cl}[B]} \int\,{\cal D}Q\; {\cal D}c\; 
            {\cal D}{\bar c}~ \exp\!\left\{ {\frac{i}{2 g_0^2}\,
              \left[ S^{\rm diam} + S^{\rm param} \right]} \right\}~,
\label{sunpartition}
\end{equation}
where 
\begin{equation} 
\begin{array}{c}
  {\ds S^{\rm diam} = - \idxd\, {\rm Tr}\, \Big\{
          \left(D_\m[B]\, Q_\n \right) \left(D^\m[B]\, Q^\n\right) 
      - 2\,{\bar c} D_\m[B]\,D^\m[B]\, c \Big\} } \\[12pt]
  {\ds S^{\rm param} = 2 \idxd\, {\rm Tr}\, \Big\{ 
      F^{\m\n}[B]\,Q^\s (S_{\m\n})_{\s\r}\, Q^\r \Big\} ~.}
\end{array}
\label{sundiaparaction}
\end{equation} 
Here $D[B]_\m=\partial_\m - i\, [B_\mu,~]$ and $F_{\m\n}[B]$ denote
the covariant derivative and the field strength for the background
field $B_\m$, the $SU(N)$ gauge field $Q_\m$ describes the quantum
fluctuations about the background and $(S_{\m\n})_{\s\r}$ are the
generators of the Lorentz group in the spin one representation, {\it
i.e.}
\begin{displaymath}
    (S_{\m\n})_{\s\r} =i \left( \e_{\m\s}\e_{\n\r} 
                              - \e_{\m\r}\e_{\n\s} \right) ~.
\end{displaymath}
The high-momentum modes contributing to the path integral $Z[B]$ 
yield a logarithmic UV divergence. This divergence shows in
dimensional regularization as a pole in $\ee = 0$, with $D=4+2\ee$.
As a matter of, introducing dimensional regularization and integrating over
$Q_\mu$, one obtains
\begin{equation} 
\begin{array}{l}
   {\ds \Gamma[B]= \frac{1}{i}\,\ln Z[B] }\\[9pt]
\hphantom{ \Gamma[B] ~}
   {\ds = -\,\frac{1}{2} \,\bigg( \frac{1}{g_0^{2}} 
                         + \frac{a^{\rm diam}}{\ee} 
                         + \frac{a^{\rm param}}{\ee} \bigg)
     \idxd\,{\rm Tr} \,\big( F^{\m\n}[B]\, F_{\m\n}[B] \big) 
     \,+\, \Gamma_{\rm finite}[B] ~,}
\end{array}
\label{UVpoles}
\end{equation}   
where\footnote{In this paper, to emphasize the value of diamagnetic
and paramagnetic contributions, we will enclose the relevant factors in
square brackets, as in eq. (\ref{coeff}).}
\begin{equation} 
    a^{\rm diam} = \frac{N}{16\pi^2}\bigg[\frac{1}{3}\bigg] \qquad
    a^{\rm param} = \frac{N}{16\pi^2}[-4]
\label{coeff}
\end{equation}
and $\Gamma_{\rm finite}[B]$ collects all finite contributions as
$D\to 4$. The origin of the coefficients $a^{\rm diam}$ and $a^{\rm
param}$ can be explained as follows. The orbital motion of the charged
quanta with only two polarizations $(D-2$ in $D$ space-time
dimensions) of the gluon field $Q_\m$ in the background $B_\m$ yields
``diamagnetism''. This motion is described by $S^{\rm diam}$ above and
its contribution to the r.h.s of eq. (\ref{UVpoles}) has a singular
part coming from high-energy quanta which is given by $a^{\rm
diam}/\ee$. On the other hand, $S^{\rm paran}$ involves the spin
current $Q^\s (S_{\m\n})_{\s\r}Q^\r$ and describes the coupling
between the spin of the gluon field $Q_\m$ and the background
$B_\m$. This coupling gives rise to ``paramagnetism'' and the
contribution to $\Gamma[B]$ of the high-momentum quanta involved in it
is $a^{\rm param}/\ee$. One next defines the renormalized coupling
constant or renormalized charge $g(\mu)$ in the MS scheme at one loop
as usual:
\begin{equation} 
    \frac{1}{g_0^{2}} + \frac{a^{\rm diam}}{\ee}
      + \frac{a^{\rm param}}{\ee} = \frac{1}{g^{2\!}(\m)\,\m^{2\ee}}~.
\label{renorcoupling}
\end{equation}
Hence, the one-loop beta function $\b(g^2)$ reads for $D=4$
\begin{displaymath} 
   \b(g^2)=\m\,\frac{d g^2}{d \m}=2\,g^{2\!}(\mu) 
       \left( a^{\rm diam} + a^{\rm param} \right) =
           \frac{N}{16\pi^2}\,g^{4\!}(\mu)\bigg[-\frac{22}{3}\bigg] ~.
\end{displaymath}
{}From this equation for the beta function one can draw the following
conclusions. First, the coefficient $a^{\rm diam}$ being positive
implies that its effect is to make the charge $g^2(\m=\Lambda)$
decrease with the momentum scale $\Lambda$. This is the charge
screening effect due to the orbital motion of the two physical
polarizations of the quanta $Q_\m$ in the field $B_\m$. Secondly,
since $a^{\rm param}$ is negative, its effect is to make
$g^2(\m=\Lambda)$ grow as $\Lambda$ decreases.  This is called
anti-screening of the charge and it is due to the interaction of the
spin with the field $B_\m$. And thirdly, the inequalities $\,|a^{\rm
param}| > a^{\rm diam}\,$ and $a^{\rm param} < 0$ explain
quantitatively the negative sign of the beta function of the theory,
hence, that the charge goes to zero as $\Lambda$ goes to infinity
(asymptotic freedom).

It is already a year since it has been shown \cite{KWMR} \cite{Jab}
that $U(1)$ gauge theory on non-commutative $\RR^4$ has an UV
divergent behaviour at one loop very similar to that of conventional
Yang-Mills theory [see \cite{Armoni} for the $U(N)$ case]. In fact,
the one-loop beta function is also negative \cite{KWMR}, which leads
to asymptotic freedom. By contrast, as discovered in ref. \cite{MRS},
the IR behaviour of non-commutative $U(1)$ gauge theory presents
completely novel features \cite{H} \cite{MST}. Indeed, the
renormalization of UV divergences induce IR divergences, thus yielding
a relation between UV and IR divergences which has been interpreted as
a sort of UV/IR duality. This duality seems not to be an artifact of
perturbation theory \cite{nonpert}, since it has been re-obtained by
defining the field theory as the infinite tension limit of the
appropriate open bosonic string theory on a magnetic $B$-field
\cite{Bfield} \cite{D3brane}.

The purpose of this paper is two-fold. First, to investigate if, in
analogy with conventional Yang-Mills theory, the one-loop beta
function of $U(1)$ gauge theory on non-commutative Minkowski space can
be understood as a dominance of paramagnetism over diamagnetism. And
secondly, to study through explicit computations UV/IR mixing from the
point of view of paramagnetism versus diamagnetism. The paper is
organized as follows. In section 2, we will explicitly compute the
two-point function and show that the one-loop beta function of $U(1)$
gauge theory on non-commutative Minkowski space has a paramagnetic
contribution, producing anti-screening of the charge, and a
diamagnetic contribution, giving rise to screening of the charge. We
will see that the paramagnetic contribution dominates, thus providing
a negative beta function. Furthermore, we will take advantage of the
computations in ref. \cite{D3brane} to show that these paramagnetic
and diamagnetic contributions can be given a stringy interpretation as
tachyon magnification and zero mode contributions, in the same sense
as for conventional Yang-Mills theory \cite{Alvarez}.  In sections, 3
and 4 we will calculate the UV divergent terms and the leading
non-commutative IR terms of the three and four-point functions.  It
will turn out that the UV/IR mixing occurs for paramagnetic and
diamagnetic logarithmic contributions separately. Section 5 collects
or conclusions. We will argue there how our results lead to an IR
renormalization of the coupling constant.

\section{The vacuum polarization tensor}

Non-commutative Minkowski space-time is defined by the algebra
generated by the operators $X^{\m}~(\m=0,\ldots,D-1)$ subject to the
commutation relations
\begin{displaymath}
   [X^{\m},X^{\n}] = i\theta^{\m\n}~,
\end{displaymath}
where $\theta^{\m\n}$ is an anti-symmetric real matrix and contraction
of indices is performed with the Minkowski metric. We shall take
$\theta^{\m\n}$ to be ``magnetic'', {\it i.e.} $\theta^{0i}=0$ for
$i=1,2,3$, since in this case the field theory exists as the zero
slope limit of a string theory in a magnetic background. If
$\theta^{\m\n}$ is ``electric'', the field theory does not exist as
the zero slope limit of a string theory \cite{SST} and does not make
sense on its own since it does not preserve unitarity
\cite{GM}. Without loss of generality, we will take $\theta^{\m\n}$ to
vanish for all $\m$ and $\n$, except for $\m,\n=1,2$, for which we
write
\begin{displaymath}
   \theta^{12}=-\theta^{21}\equiv\theta~.
\end{displaymath}  

The classical action of $U(1)$ gauge theory on non-commutative
Minkowski space-time is given by
\begin{equation}
   S_{\rm class}[A] = -\frac{1}{4 g^2}\idxd 
         \left(\,F^{\m\n}\star F_{\m\n}\,\right)(x)~,
\label{classaction}
\end{equation}
where
\begin{displaymath}
\begin{array}{c}
     F_{\m\n}(x)= \partial_{\m}A_\n (x) - \partial_{\n}A_\m (x) 
                - i\, [A_\m,A_\n](x)\\[12pt]
     [A_\m,A_\n] = (A_\m\star A_\n-A_\n\star A_\m )(x) 
\end{array}
\end{displaymath}
is the field strength and the symbol $\star$ stands for the Moyal
product of functions
\begin{displaymath}
   \big(f\star g\big) (x) = f(x)\, 
         e^{\frac{i}{2}\,\theta^{\m\n}\, \overleftarrow {\partial_\m}
             \, \overrightarrow{\partial_\n}}\, g(x)~.
\end{displaymath}
The classical theory is invariant under non-commutative $U(1)$ gauge
transformations, which in infinitesimal form have the form
\begin{displaymath}
   \delta_{\om} A_{\m}(x)=D_\m[A]\,\om = 
            \partial_\m\om(x)-i\,[A_\m,\om](x)~,
\end{displaymath}
the commutator being defined with regard to the Moyal product.

To quantize the theory around a background field configuration, say
$B_\m(x)$, we shall use the background field method in the Feynman-`t
Hooft gauge. To this end, we write the gauge field $A_\m$ as the sum
of the the background $B_{\m}$ and the quantum fluctuation $Q_\m$
about it, $A_\m \!= B_\m\!+Q_\m$. The tree-level action $S$ then
becomes
\begin{equation}
      S = S_{\rm class}[B+Q] + S_{\rm gf} + S_{\rm gh},
\end{equation} 
where the gauge fixing and ghost terms have the form
\begin{displaymath}
\begin{array}{c}
    {\ds S_{\rm gf} = -\frac{1}{2}\, \idxd\, 
         D_\m[B]\, Q^\m\! \star D_\n[B]\,Q^\n } \\[12pt]
    {\ds S_{\rm gh} = \idxd\,\bar{c} \star 
         D_\m[B\!+\!Q]\, D^\m[B]\,c ~.}
\end{array}
\end{displaymath}
The fields $c$ and $\bar c$ are the ghost fields and the commutators
in the covariant derivatives
\begin{displaymath}
     D_\m[B\!+\!Q]= \partial_\m -i\,[B_\m+Q_\m,\;\,] \qquad 
     D_\m(B)=\partial_\m -i\,[B_\m,\;\,]
\end{displaymath}
are defined with regard to the Moyal product. Some straightforward
manipulations give for the partition function $Z[B]$ of the theory up
to one loop in the $U(1)$ background field $B_\m$ the expression
\begin{equation} 
      Z[B]= e^{iS_{\rm cl}[B]} \int\,{\cal D}Q\; {\cal D}c\; 
            {\cal D}{\bar c}~ \exp\!\left[ \frac{i}{2 g_0^2}\,
              \left( S^{\rm diam} + S^{\rm param} \right) \right]~,
\label{uonepartition}
\end{equation}
with 
\begin{equation} 
\begin{array}{c}
  {\ds S^{\rm diam} = -\idxd\,\Big( D_\m[B]\, Q_\n \star D^\m[B]\, Q^\n 
      - 2\,{\bar c}\star D_\m[B]\,D^\m[B]\, c \Big)} \\[12pt]
  {\ds S^{\rm param} = 2 \idxd\, F^{\m\n}[B]\star\,Q^\s\star 
      (S_{\m\n})_{\s\r}\, Q^\r ~,}
\end{array}
\label{diaparaction}
\end{equation} 
in complete analogy with eq. (\ref{sunpartition}).  The term $S^{\rm
param}$ involves the spin one non-commutative current $Q^\s\star
(S_{\m\n})_{\s\r}\,Q^\r$ and describes the coupling of the spin to the
background field $B_\m$. This term thus gives rise to non-commutative
Pauli ``paramagnetism''.  In turn, the functional $S^{\rm diam}$ is
the classical action for the motion in the field $B_\m$ of the $D-2$
physical degrees of freedom of the field $Q_\m$. Indeed,
\begin{equation}
  \int {\cal D}Q\;{\cal D}c\; {\cal D}{\bar c} ~ 
     \exp\!\left[ \frac{i}{2 g_0^2}\, S^{\rm diam} \right] = 
       \left( {\rm det}^{-1/2}\,D^2[B] \,\right)^{D-2}.
\label{Dminus2}
\end{equation} 
We shall then say that $S^{\rm diam}$ gives rise to non-commutative
Landau ``diamagnetism''. From $S^{\rm diam}$ and $S^{\rm param}$ one
readily extracts the Feynman rules needed for one-loop perturbative
computations. We have collected them in fig. 1, where we have used the
notation 
\begin{displaymath}
    q\wedge p=\theta^{\m\n}q_\m p_\n ~.
\end{displaymath}
Vertices coming from $S^{\rm diam}$ and $S^{\rm param}$ will be called
diamagnetic and paramagnetic respectively.

In this section we compute up to one loop the vacuum polarization
tensor $\Pi_{\m\n}(p)$, defined as 
\begin{displaymath}
   i\Pi_{\m\n}(x,y) = \frac{\d^2 \Gamma[B]}
          {\d B_\m(x)\d B_\n(y)}{\bigg\arrowvert}_{B=0} 
       =  \idpd\, e^{-ip\cdot(x-y)}\,i\Pi_{\m\n}(p) ~,
\end{displaymath} 
where $i\Gamma[B]=\ln Z[B]$. According to the nature of their
vertices, the one-loop Feynman diagrams contributing to
$\Pi_{\m\n}(p)$ fall into three categories: diagrams with only
diamagnetic vertices, diagrams with only paramagnetic vertices and
diagrams with both diamagnetic and paramagnetic vertices. The
contributions to the vacuum polarization tensor coming from these
three categories will be denoted by $\Pi_{\m\n}^{\rm diam}(p)$,
$\Pi_{\m\n}^{\rm param}(p)$ and $\Pi_{\m\n}^{\rm mixed}(p)$, so we
write
\begin{equation}
   i\Pi_{\m\n}(p) = i\Pi_{\m\n}^{\rm diam}(p)
                  + i\Pi_{\m\n}^{\rm param}(p) 
                  + i\Pi_{\m\n}^{\rm mixed}(p) ~.
\label{poltensor}
\end{equation} 
It is very easy to see that there is only one diagram contributing to
$i\Pi_{\m\n}^{\rm mixed}(p)$, namely that depicted in fig 2. Upon
performing some algebra in the integrand, the corresponding Feynman
integral takes the form
\begin{displaymath}
    2i\idqd\, (\eta_{\m\n} - \eta_{\m\n})\,
        \sin^2\left(\frac{q\wedge p}{2}\right)\,
            \frac{(2q+p)_\r}{q^2(q+p)^2} = 0 ~.
\end{displaymath}
Hence
\begin{equation}
            \Pi_{\m\n}^{\rm mixed}(p) = 0~.
\label{polmixed}
\end{equation}
We are thus left with the diamagnetic and paramagnetic
contributions. 

The one-loop contribution to $i\Pi_{\m\n}^{\rm diam}(p)$ is given by the
sum of the diagrams in fig. 3, which using the Feynman rules reads
\begin{equation}
  i\Pi_{\mu\nu}^{\rm diam}(p) = [D-2] \idqd\, 2\, 
      \sin^2\!\left(\frac{q\wedge p}{2}\right) 
      \left[ \frac{(p+2q)_\m\, (p+2q)_\n}{q^2\,(p+q)^2}
           - 2\,\frac{\eta_{\mu\nu}}{q^2}\right]~.
\label{poldiam}
\end{equation}
The $D$ in the factor $[D-2]$ in front of the integral comes from the
diagrams with photons flowing around the loop, whereas the diagrams
with ghost loops yield the contribution $-2$.  Hence the effective
one-loop contribution to $i\Pi_{\mu\nu}^{\rm diam}(p)$ corresponds to
$D-2$ scalar fields (transforming in the adjoint representation of the
gauge group) moving in the field $B_\m$, in agreement with the
discussion following eqs. (\ref{uonepartition}) and
(\ref{diaparaction}) above. To keep to a minimum the number of
diagrams to draw, we have not considered planar and non-planar
diagrams separately. In fact, each of our diagrams is the sum of a
planar and a non-planar contribution. To disentangle these
contributions from one another, it is enough to use the identity $\,4
\sin^2\!\left(q\wedge p/2\right) = 2 - e^{iq\wedge p/2}- e^
{-iq\wedge p/2}$. This identity gives, upon substitution in
(\ref{poldiam}), a $\theta^{\m\n}\!$-independent integral, which
defines the planar contribution $i\Pi^{\rm diam,P}(p)$, and a
$\theta^{\m\n}\!$-dependent integral, which constitutes the non-planar
contribution $i\Pi^{\rm diam,NP}(p)$:
\begin{equation}
  i\Pi_{\mu\nu}^{\rm diam}(p) = i\Pi_{\mu\nu}^{\rm diam,P}(p) 
      + i\Pi_{\m\n}^{\rm diam,NP}(p) ~.
\end{equation}
After introducing Schwinger parameters, performing the momenta
integrals in dimensional regularization and integrating by parts to
factorize out the transverse tensor $p^2\eta_{\m\n}\!-p_\m p_\n$, we
obtain
\begin{equation}
   i\Pi_{\mu\nu}^{\rm diam,P}(p) = -i\,\frac{[D-2]}{(4\pi)^{D/2}}\;
      (p^2\eta_{\mu\n}-p_\m p_\n) \int_0^1\! dx\; (1-2x)^2
         \int_0^{\infty}\! dt\; t^{1-D/2} \;e^{- p^2 t x(1-x)}       
\label{plandiam}
\end{equation}
and
\begin{equation}
\begin{array}{rl}
   {\ds i\Pi_{\mu\nu}^{\rm diam,NP}(p) = i\;\frac{[D-2]}{(4\pi)^{D/2}} }
                  \hspace{-6pt} 
     & {\ds \bigg[\,(p^2\eta_{\m\n}-p_\m p_\n) \int_0^1\! dx\;(1-2x)^2  
          \int_0^\infty\! dt\; t^{1-D/2}\; 
                e^{- p^2 t x(1-x)-\,\tp^2/4t} } \\[12pt]
      & + \;{\ds \tp_\m  \tp_\n \int_0^1\! dx 
          \int_0^\infty\! dt\; t^{-1-D/2}\; 
                e^{ - p^2 tx(1-x) - \,\tp^2/4t}\,\bigg] }~,
\end{array}
\label{nonplandiam}
\end{equation}
where 
\begin{displaymath}
    {\tilde p}^\m =  \theta^{\m\n} \, p_\n \qquad
    {\tilde p}^2=p_\m\, \theta^{\m\r}\,\theta_\r{}^\n\, p_\n ~.
\end{displaymath}
It is well known \cite{Speer} that when a Feynman amplitude is
expressed in terms of Schwinger parameters (the so-called parametric
representation) the contribution to the amplitude coming from virtual
quanta carrying arbitrarily high momenta is given by the contribution
to the corresponding parametric integral coming from regions of the
integration domain where all the Schwinger parameters are arbitrarily
small. If we apply this reasoning to $i\Pi_{\mu\nu}^{\rm diam,P}(p)$,
we conclude that, for $D\geq 4$, the non-integrable singularity in
eq. (\ref{plandiam}) at $t=0$ shows that virtual quanta $Q_\m$
carrying arbitrarily high momenta yield an UV divergent
contribution. This divergence needs renormalization. On the other
hand, if ${\tilde p}^2 \neq 0$, the integrand of any of the integrals
in eq. (\ref{nonplandiam}) is non-singular at $t=0$. Hence the
contribution to $i\Pi_{\mu\nu}^{\rm diam,NP}(p)$ coming from arbitrary
high momenta quanta $Q_\m$ is curbed by the non-commutativity of space
through the exponential $\exp(-\tp^2/4t)$, acting
$1/\tp^2$ as a regulator. Of course, if the regulator is
removed, divergences spring back. This explains the UV origin of the
IR divergences that occur in $i\Pi_{\mu\nu}^{\rm diam,NP}(p)$ at
$\tp^2=0$. Furthermore, as it will come up shortly, the fact
that the tree-level vertices vanish as $\theta^{\m\n}\to 0$, so that
loops formally vanish in this limit, makes the coefficient of the
logarithmic UV divergence in $i\Pi_{\mu\nu}^{\rm diam,P}(p)$ at $D=4$
to be opposite to the coefficient of its logarithmic IR divergence at
${\tilde p}^2 = 0$. All this is the UV/IR mixing at work
\cite{MRS}.

The one-loop paramagnetic contribution $i\Pi_{\m\n}^{\rm param}(p)$
to the vacuum polarization tensor is given by the contribution
involving only two background fields which comes from the diagram in
fig. 4.  For the diagram itself the Feynman rules yield
\begin{equation}
   {\left< F_{\m\n}(-p) F_{\r\s}(p) \right>}_{\rm 1PI} =
     4\, (\eta_{\m\r}\eta_{\n\s} -\eta_{\m\s}\eta_{\n\r})\, J(p) ~,
\label{FFparam}
\end{equation}
where $J(p)$ is the integral
\begin{equation}
   J(p) = \idqd\, 2\sin^2\left(\frac{q\wedge p}{2}\right)\, 
           \frac{1}{q^2\, (p+q)^2}~.
\label{Jofp}
\end{equation}
The contribution $i\Pi_{\m\n}^{\rm param}(p)$ is then given by 
\begin{equation}
   i\Pi_{\m\n}^{\rm param}(p) = [\,8\,]\,i\, 
       (p^2\,\eta_{\m\n}-p_\m p_\n)\,J(p)~,
\label{polparam}
\end{equation}
which has a planar and a non-planar part
\begin{displaymath}
   i\Pi_{\m\n}^{\rm param}(p) = i\Pi_{\m\n}^{\rm param,P}(p) 
                              + i\Pi_{\m\n}^{\rm param,NP}(p)~.
\end{displaymath}
Using the same arguments as for the diamagnetic part, we obtain
\begin{equation}
\begin{array}{c}
   i\Pi_{\m\n}^{\rm param,P}(p) 
      = [\,8\,]\,i\,(p^2\,\eta_{\m\n}-p_\m p_\n)\, J^{\rm P}(p)\\[12pt] 
   {\ds    J^{\rm P}(p) = \frac{1}{(4\pi)^{D/2}}\; \int_0^1\! dx\, 
        \int_0^\infty\! dt \; t^{1-D/2} e^{- p^2 t x(1-x)} }
\end{array}
\label{planparam}
\end{equation}
for the planar part, and
\begin{equation}
\begin{array}{c}
   i\Pi_{\m\n}^{\rm param,NP}(p) 
      = [\,8\,]\,i\,(p^2\,\eta_{\m\n}-p_\m p_\n)\, J^{\rm NP}(p) \\[12pt]
  {\ds J^{\rm NP}(p) =  -\,\frac{1}{(4\pi)^{D/2}} \;
       \int_0^1\! dx \int_0^\infty\! dt\;
       t^{1-D/2} e^{ - p^2 t x(1-x) - \,\tp^2/4t } }
\end{array}
\label{nonplanparam}
\end{equation}
for the non-planar part. A similar analysis as for the diamagnetic
contributions is in order. Indeed, the high momenta modes going around
the loop of the paramagnetic diagram in fig. 4 give, for $D\geq 4$, an
UV divergent contribution to $J^{\rm P}(p)$, hence to $i\Pi_{\m\n}^{\rm
param,P}(p)$, which corresponds to the non-integrable singularity at
$t=0$ in eq. (\ref{planparam}). The contribution of these modes to
$J^{\rm NP}(p)$, hence to $i\Pi_{\mu\nu}^{\rm NP}(p)$ is finite
provided ${\tilde p}^2\neq 0$. We may then say that the
non-commutative character of space makes $1/\tp^2$ to play the r\^ole
of an UV regulator for the high-momentum modes contributing to the
non-planar part of $i\Pi_{\mu\nu}^{\rm param}(p)$. If we remove the
regulator, {\it i.e.}  take the limit $1/\tp^2\to \infty$ ($\tp^2\to
0$), the divergence is recovered, although this time under the guise
of an IR divergence.

We are now ready to compute the contributions to the one-loop beta
function of the theory at $D=4$ coming from the diamagnetic and
paramagnetic parts of the vacuum polarization tensor. Recall that the
mixed part vanishes. If we set $D=4+2\ee$ in eqs. (\ref{plandiam}) and
(\ref{planparam}) and make a Laurent expansion around $\ee=0$, we
obtain
\begin{equation}
\begin{array}{c}
  {\ds i\Pi_{\m\n}^{\rm diam,P}(p) = i\; (p^2\eta_{\m\n}-p_\m p_\n)\, 
     \left[ \frac{a^{\rm diam}}{\ee} 
          + a^{\rm diam}\,\ln\!\left(\!-\frac{p^2}{4\pi}\right) 
          - \frac{5}{72\pi^2} + O(\ee) \right]} \\[12pt]
  {\ds i\Pi_{\m\n}^{\rm param,P}(p) = i\;(p^2\eta_{\mu\n}-p_\m p_\n)\,
     \left[ \frac{a^{\rm param}}{\ee} 
          +a^{\rm param}\,\ln\!\left(\!-\frac{p^2}{4\pi}\right) 
          + \frac{1}{\pi^2} + O(\ee) \right] } ~,
\end{array}
\label{Laurent}
\end{equation}
where
\begin{equation} 
     a^{\rm diam} = \frac{1}{16\pi^2}\; \bigg[\frac{2}{3}\bigg] \qquad
     a^{\rm param} = \frac{1}{16\pi^2}\; [-8]~.
\label{coeffnon}
\end{equation}
Note that the coefficients $a^{\rm diam}$ and $a^{\rm param}$ can be
obtained from the corresponding coefficients in eq. (\ref{coeff}) by
replacing in the latter $N$ with $2$. To subtract UV divergences, we
work in the MS renormalization scheme and add to the classical action
$S_{\rm class}[B]$ the counterterm
\begin{equation}
   \d S[B] = \frac{1}{4\ee} \left( a^{\rm diam}+a^{\rm param} \right) 
             \idxd\, F^{\m\n}[B] \star F_{\m\n}[B] ~,
\label{counterterm}
\end{equation} 
since its term quadratic in the background field $B_\m$ cancels the UV
divergences in eqs. (\ref{Laurent}).
In sections 3 and 4 we will see that the UV divergences in the three
and four-point function are also subtracted by the counterterm 
(\ref{counterterm}). Taking now into account that the
tree-level part of the vacuum polarization tensor in terms of the bare
coupling constant $g_0^2$ reads
\begin{displaymath}
    -\,\frac{i}{g_0^2}\;(p^2\eta_{\mu\n}-p_\m p_\n)
\end{displaymath}
and using eqs. (\ref{Laurent}), we conclude that up to one loop the
renormalized vacuum polarization tensor in the MS scheme takes the
form
\begin{equation}
\begin{array}{l}
   {\ds i\Pi_{\m\n}^{\rm ren}(p) = i\; (p^2\eta_{\m\n}-p_\m p_\n)\; 
      \left[ -\,\frac{1}{g^2} 
          + \left( a^{\rm diamag}+a^{\rm param}\right)\, 
                  \ln\!\left(\!- \frac{p^2}{4\pi\m^2}\right) 
          +\frac{67}{72\pi^2} \right] } \\[12pt]
   {\ds \phantom{\Pi_{\m\n}^{\rm ren)}(p)~} 
     + i\Pi_{\m\n}^{\rm diam,NP}(p) 
     + i\Pi_{\m\n}^{\rm param,NP}(p) ~,}
\end{array}
\label{renvacpol}
\end{equation}
where the non-planar contributions $i\Pi_{\mu\nu}^{\rm diam,NP}(p)$ and
$i\Pi_{\mu\nu}^{\rm param,NP}(p)$ are obtained by setting $D=4$ in
eqs. (\ref{nonplandiam}) and (\ref{nonplanparam}) respectively. The
letter $g$ in eq. (\ref{renvacpol}) is the renormalized coupling
constant in the MS scheme, given by eq. (\ref{renorcoupling})
for $a^{\rm diam}$ and $a^{\rm param}$ as in eq. (\ref{coeffnon}). The
beta function of the theory then reads
\begin{equation} 
   \b(g^2) = \m\;\frac{d g^2}{d \m}= 2\,g^{2\!}(\mu)
     \left( a^{\rm diam} + a^{\rm param} \right) = \frac{1}{4\pi^2}\; 
          g^{4\!}(\mu) \left[-\,\frac{11}{3}\right]~.
\label{nonbeta}
\end{equation}
We see that the negative sign of the beta function comes about because
the high-momentum paramagnetic contributions, which yield the
coefficient $a^{\rm param}$, overcome the high-momentum diamagnetic
contributions, which originate the coefficient $a^{\rm diam}$. The
analogy with $SU(N)$ theory on commutative Minkowski space-time is
clear.

Let us next show that the dominance at high energies of paramagnetic
contributions over diamagnetic ones has, due to the UV/IR mixing, a
cut effect on the IR behaviour\footnote{The IR divergences that occur
when certain linear combinations $l_i (p)$ of the external momenta
$p_j$ satisfy $\tilde{l_i}=0$, with $l_i\neq 0$, will be called
non-commutative IR divergences throughout this paper.  The behaviour
of the Green functions in the neibourhood of those momentum
configurations will be called non-commutative IR behaviour.}  of the
theory at $\tp=0$. We start by noting that the divergent IR behaviour
at $\tp=0$ of the renormalized vacuum polarization tensor in
eq. (\ref{renvacpol}) is only caused by $i\Pi_{\mu\nu}^{\rm diam)\,
NP}(p)$ and $i\Pi_{\mu\nu}^{\rm param)\, NP}(p)$. The $\tp=0$ divergent
terms of the latter are easily computed from eqs. (\ref{nonplandiam})
and (\ref{nonplanparam}). For them we obtain
\begin{equation}
\begin{array}{c} 
   i\Pi_{\mu\nu}^{\rm diam,NP}(p)\, \approx\, 
       - i\,a^{\rm diam}\, \ln(-p^2\,\tp^2)\; 
             (p^2\eta_{\mu\n}-p_\m p_\n) 
       + {\ds \frac{2i}{\pi^2}\; 
           \frac{\tp_\m\tp_\n}{(\tp^2)^2}} \\[12pt]
   i\Pi_{\mu\nu}^{\rm param,NP}(p)\, \approx\, 
       - i\,a^{\rm param}\, \ln(-p^2\,\tp^2)\; 
              (p^2\eta_{\mu\n}-p_\m p_\n) ~,
\end{array}
\label{IRterms}
\end{equation}
where $\approx$ means that we have dropped any term which is finite at
$D=4$ as $\tp\rightarrow 0$. It is plain that the IR
divergent logarithmic contributions in eqs. (\ref{IRterms}) are dual
to the UV divergent logarithmic terms in eqs. (\ref{Laurent}), in the
sense that in the momenta region
\begin{equation}
   |\tilde p| \sim \theta \Lambda_{\rm IR}
\label{2plog}
\end{equation}
the former can be obtained from the latter by using the replacements 
\begin{equation}
     \frac{1}{\ee} \leftrightarrow \ln \La_{\rm UV}^2 \qquad 
     \La_{\rm UV} \leftrightarrow \frac{1}{\theta\Lambda_{\rm IR}} ~.
\label{identifications}
\end{equation}
By contrast, the quadratic IR divergent term $\tp_\m\tp_\n/(\tp^2)^2$
in eq. (\ref{IRterms}) has no dual counterpart as a singular UV
contribution in eq. (\ref{Laurent}). This is due to the fact that, if
gauge invariance is preserved, no local and quadratic UV divergent
contribution occurs in the vacuum polarization tensor. Yet, as can be
seen from eq. (\ref{nonplandiam}), the origin of both logarithmic and
quadratic divergences at $\tilde p=0$ is the same: the curbing by the
non-commutative character of the space, through the exponential
$\exp(-\tp^2/4t)$, of the non-planar contribution coming from the
high-momentum quanta flowing along the loop of the diagrams in
fig. 3. Note that both the diamagnetic and paramagnetic functionals
$S^{\rm diam}$ and $S^{\rm param}$ in eq. (\ref{diaparaction})
contribute to the non-commutative IR logarithmic divergence of the
vacuum polarization tensor, whereas the quadratic divergence only
receives contributions from the diamagnetic functional, describing the
orbital motion of $D-2$ scalar quanta in the field $B_\m$. We shall
see in the next section that the non-logarithmic IR divergences at
$\tp=0$ of the one-loop three-point function also have a purely
diamagnetic origin. Finally, adding the contributions in
eq. (\ref{IRterms}), we conclude that
\begin{equation} 
   i\Pi_{\m\n}(p)\,\approx\, \frac{i}{16\pi^2}\,
       \left[ \frac{22}{3} \right]\, \ln(-p^2\,\tp^2)\,
              (p^2\eta_{\m\n}-p_\m p_\n) 
   + \frac{2i}{\pi^2}\; \frac{\tilde p_\m\tilde p_\n}{\tilde p^4}
   \equiv i\Pi_{\m\n}^{\rm IR}(p) ~.
\label{vacuumIR}
\end{equation}
We have thus shown that the phenomenon of paramagnetic dominance at
high energies, $|a^{\rm param}| > a^{\rm diam}$, which explains
the sign of the one-loop beta function, also renders, via UV/IR
mixing, the coefficient of the logarithmic divergence
$\ln(-p^2\,\tp^2)$ in the vacuum polarization tensor equal to
$22/3$. This coefficient and the coefficient of $\ln(-p^2/4\pi\m^2)$
in eq. (\ref{renvacpol}) have the same absolute value but opposite
sign, yet another manifestation of the mixing.

We shall close this section by showing that the right hand side of
eqs. (\ref{plandiam}), (\ref{nonplandiam}), (\ref{planparam}) and
(\ref{nonplanparam}) also arise in the infinite tension limit of open
string theory in a constant magnetic field.  We will follow
ref. \cite{D3brane} and obtain the non-commutative $U(1)$ theory as
the infinite tension limit of an open bosonic string theory on a
D3-brane stuck at an $\RR^{22}/Z_2$ orbifold singularity with a
constant magnetic field along the worldvolume directions of the
brane. The reader is referred to ref. \cite{D3brane} for a
comprehensive discussion.  Let us first compute the one-loop
non-planar contribution to the string amplitude with two photon vertex
insertions of polarizations $e^{(1)}_\m$ and $e^{(2)}_\n$. This is
given by
\begin{displaymath}
   {\cal A}^{\rm NP} = e^{(1)}_\m\, i\Pi^{\rm NP}{}^{\m\n}\,e^{(2)}_\n~,
\end{displaymath}
where
\begin{equation}
\begin{array}{l}
  {\ds i\Pi^{\rm NP}{}^{\m\n} = g^2\! \int_{0}^\infty\!\frac{dt}{2t} 
    \int_{0}^{2\pi t}\!dy_1 \int_{0}^{2\pi t}\!dy_2 \;Z(t) }\\[12pt]
\hphantom{i\Pi^{\rm NP}{}^{\m\n} ~}
   {\ds \times\, \left(\!-\,p_\r p_\s \,
                          \partial_{y_1} \tilde{\cal G}^{\r\s}\,
                          \partial_{y_2} \tilde{\cal G}^{\m\n} 
        + p_\r p_\s \, \partial_{y_1} \tilde{\cal G}^{\m\r}\,
                        \partial_{y_2} \tilde{\cal G}^{\n\s}\right)
   e^{p_\r p_\s \tilde{\cal G}^{\rho\sigma}} ~,}
\end{array}
\label{stringtwopoint}
\end{equation}
$Z(t)$ is the open string partition function of an open bosonic
string ending on a D$p$-brane glued at an orbifold singularity in the
presence of a constant magnetic field along the directions of the
brane, and $\tilde{\cal G}^{\m\n}$ is related to the string propagator
and reads
\begin{equation}
   \tilde{\cal G}^{\m\n}= - \alpha'\, 
      \bigg[\, G^{\m\n}\,\gm^{\rm NP}(y|t) 
           + \frac{i\,\theta^{\m\n}}{2\pi\a'}\;\frac{y}{t}
           + \frac{(\theta G\theta)^{\m\n}}{(2\pi\alpha')^2} \;
             \frac{\pi}{2t}\,\bigg]~.
\label{tpropagator}
\end{equation}
Here $G^{\m\n}$ is the inverse of the effective open string metric
$G_{\m\n}$ \cite{SW}, $y=y_1-y_2$ and $\gm^{\rm NP}$ has the form
\begin{equation}
    \gm^{\rm NP}(y|t) = \ln\!\Bigg\arrowvert 2\pi\,
         \frac{\vartheta_2(iy/2\pi,it)}{\vartheta_1'(0,it)}
                        \Bigg\arrowvert^2 -\, \frac{y^2}{2\pi t} ~.
\label{Gammanp}
\end{equation}
From eqs. (\ref{tpropagator}) and (\ref{stringtwopoint}), one readily
has
\begin{equation}
\begin{array}{l}
  {\ds i\Pi^{\rm NP}{}^{\m\n}= g^2\!\int_0^\infty \frac{dt}{2t}
         \int_0^{2\pi t}\!dy_1 \int_0^{2\pi t}\!dy_2 \; Z(t)\;{\a'}^2\, 
         \bigg\{ (p^2 G^{\m\n} - p^\m p^\n)\, 
                 \Big[\partial_y \gm^{\rm NP}(y|t)\Big]^2 }\\[15pt]
\hphantom{i\Pi^{\rm NP}{}^{\m\n} ~ }
   {\ds +\, \frac{i\,(p^2\theta^{\m\n} -p^\m\tp^\n -p^\n\tp^\m)}
                 {2\pi\alpha'}\; \frac{1}{t}\;
            \partial_y\gm^{\rm NP}(y|t)
         +\frac{\,\tp^\m\tp^\n}{(2\pi\alpha')^2}\; \frac{1}{t^2} 
         \bigg\}\; e^{-\a' p^2\,\gm^{\rm NP}(y|t)}\;
                   e^{-\tp^2/8\pi\a't}~,}
\end{array}
\label{twopointeq}
\end{equation}
with $\,p^2=G_{\m\n}p^\m p^\n$, $p_\n=G_{\n\m} p^\m$ and
$\tp^\m=\theta^{\m\n}p_\n$. To obtain from this expression the
corresponding field theory result, one needs the large $t$ expansion
of the functions in the integrand upon performing the change of
variables $y=2\pi x t$. The relevant terms of these expansions are
\begin{eqnarray}
    & {\ds Z(t)\sim \frac{V_d}{(8\pi^2 \alpha' t)^{d/2}}\;
              \left( e^{2\pi t} + d-2 \right) } & 
\label{Zasymptotics}\\[9pt]
    & {\ds \gm^{\rm NP}(2\pi x t|t) \sim 2\pi tx\,(1-x) } &
\label{Gammaasympt}\\[9pt]
    & {\ds \partial_y \Gamma^{\rm NP}(y|t)\bigg|_{y=2\pi xt} 
           \sim 1-2x+2 \left(e^{-2\pi t}e^{2\pi xt}-e^{-2\pi xt} 
                        \right) ~,} &
\label{partGamasympt}
\end{eqnarray}  
where we have taken a ${\rm D}(d-1)$-brane to compute
$Z(t)$. Substituting these expansions in eq. (\ref{twopointeq}) and
making the changes of scale $t\to t/\a'$ and $y\rightarrow y/\a'$, one
obtains after some calculations\footnote{This includes dropping
divergent contributions as $t\to\infty$ for arbitrary $x$.} the
non-planar contribution to the vacuum polarization tensor as the sum
\begin{displaymath}
   i\Pi^{\rm NP}_{\m\n}(p) = i\Pi^{\rm NP,\,tach}_{\m\n}(p)
            + i\Pi^{\rm NP\,,\,0-mode}_{\m\n}(p)
\end{displaymath}
of two terms: 
\begin{equation}
\begin{array}{rl}
   {\ds i\Pi_{\mu\nu}^{\rm NP,\,0-mode}(p) = ig_0^2\;
         \frac{[d-2]}{(4\pi)^{d/2}} }\hspace{-6pt} 
     & {\ds \bigg[\,(p^2 G_{\m\n}-p_\m p_\n) \int_0^1\! dx\;(1-2x)^2  
          \int_0^\infty\! dt\; t^{1-d/2}\; 
                e^{- p^2 t x(1-x)-\,\tp^2/4t} } \\[12pt]
      & + \;{\ds \tp_\m  \tp_\n \int_0^1\! dx 
          \int_0^\infty\! dt\; t^{-1-d/2}\; 
                e^{ - p^2 tx(1-x) - \,\tp^2/4t}\,\bigg] }
\end{array}
\label{zeromode}
\end{equation}
and
\begin{equation}
   i\Pi^{\rm NP,\,tach}_{\m\n}(p) = -\,i g^2_0\;
     \frac{[-\,8\,]}{(4\pi)^{d/2}} \left(p^2 G_{\mu\n} - p_\m p_\n\right) 
      \int_0^1 dx \int_0^\infty  dt\;
       t^{1-d/2}\, e^{- p^2 t x(1-x)\, - \tp^2/4t} ~. 
\label{t.magnification}
\end{equation}
The same arguments as those used in ref. \cite{Alvarez} for Yang-Mills
theory on commutative Minkowski space-time show that
$i\Pi_{\mu\nu}^{\rm NP,\,0-mode}(p)$ originates from the product of the
one-loop photon contribution to the partition function and the
zero-mode contribution to the string Green function. In turn,
$i\Pi_{\mu\nu}^{\rm NP,\,tach}(p)$ arises from the combination of the
tachyonic contribution to the partition function and the exponentially
vanishing part of the string Green function, an effect called tachyon
magnification in ref. \cite{Alvarez}.
This can be understood as follows.

Indeed, the contribution proportional to $p^2 G_{\m\n}-p_\m p_\n$ in
$i\Pi_{\mu\nu}^{\rm NP,\,0-mode}(p)$ arises from combining the term
$d-2$ in eq. (\ref{Zasymptotics}), carrying the one-loop photon
contribution to $Z(t)$, with the square of the zero-mode term $1-2x$
in eq. (\ref{partGamasympt}). And the term proportional to $\tp_\m
\tp_\n$ in $i\Pi_{\mu\nu}^{\rm NP,\,0-mode}(p)$, note that it comes
with the extra zero-mode factor $1/t^2$. As concerns
$i\Pi_{\mu\nu}^{\rm NP,\,tach}(p)$ , we note that it
originates from the term in eq. (\ref{twopointeq}) proportional to
$p^2 G_{\m\n}-p_\m p_\n$ that goes with the product
\begin{displaymath}
   e^{2\pi t} \times 
      4 \left( e^{-2\pi t}\,e^{2\pi xt}-e^{-2\pi xt}\right)^2 ~.
\end{displaymath}
The first factor in this product, $e^{2\pi t}$, is supplied by the
large-$t$ expansion of $Z(t)$ and is due to the open string
tachyon. The second factor arises from squaring the term $2(e^{-2\pi
t}\,e^{2\pi xt} - e^{-2\pi xt})$ in eq.  (\ref{partGamasympt}) and has
its origin in the open string propagator. It is precisely $-8 e^{2\pi
t}$, the evanescent large-$t$ and $x$-independent contribution to this
second factor, which is magnified when combined with the $t$-divergent
tachyon contribution to $Z(t)$, all in all yielding a finite non-zero
result.  Note that eqs. (\ref{zeromode}) and (\ref{t.magnification})
agree with eqs. (\ref{nonplandiam}) and (\ref{nonplanparam}), if we
identify $G_{\m\n}$ with $\eta_{\m\n}$ and $d$ with $D$. Hence we have
identified the diamagnetic and paramagnetic non-planar contributions
in field theory with the zero-mode and tachyon magnification
contributions in string theory, an observation first made for
Yang-Mills theory in commutative Minkowski space time in
ref. \cite{Alvarez}.

A similar analysis can be performed for the planar contribution to the
string amplitude with two photon vertex insertions. It is this
contribution that develops an UV divergence when the field theory
limit is taken. Furthermore, in this limit, this contribution can be
decomposed in the sum of a zero-mode part and a tachyon magnification
part, whose expressions agree with the r.h.s. of eqs. (\ref{plandiam})
and (\ref{planparam}) respectively.  Thus, the charge screening
contribution to the beta function --given by the coefficient
$a_1^{diam\!}$-- is concocted by the diamagnetic contributions in the
field theory setting and has a zero mode origin in string theory
framework. On the other hand, the charge anti-screening contribution
--provided by $a_1^{\rm param\!}$-- has a paramagnetic origin in field
theory and is the result of tachyon magnification in string theory.

\section{The three-point function}

In this section we study, in the light of high-energy paramagnetic
dominance, the UV and non-commutative IR divergences of the
three-point function and their mixing. The 1PI three-point function
$\Gamma_{\m_1\m_2\m_3}(x_1,x_2,x_3)$ is defined as
\begin{displaymath}
   i\Gamma_{\m_1\m_2\m_3}(x_1,x_2,x_3) = \frac{\d^3 \Gamma[B]}
      {\d B_{\m_1}(x_1)\,\d B_{\m_2}(x_2)\, \d B_{\m_3}(x_3)}
          {\bigg\arrowvert}_{B=0} ~,
\end{displaymath}        
where $i\Gamma[B]=\ln Z[B]$ and $Z[B]$ is given in eq.
(\ref{uonepartition}). In momentum space we will write $i\gm_{\m_1 \m_2
\m_3}(p_1,p_2,p_3)$, with $p_1+p_2+p_3=0$.
The three-point function can be expressed as the sum of a diamagnetic
contribution, which sums over diagrams with all vertices of
diamagnetic type, a paramagnetic contribution, given by diagrams with
only paramagnetic vertices, and a mixed contribution, made of diagrams
with at least one vertex of each type:
\begin{displaymath}
   i\gm_{\m_1 \m_2 \m_3}(p_1,p_2,p_3) 
         = i\gm_{\m_1 \m_2 \m_3}^{\rm diam}(p_1,p_2,p_3)
         + i\gm_{\m_1 \m_2 \m_3}^{\rm param}(p_1,p_2,p_3)
         + i\gm_{\m_1 \m_2 \m_3}^{\rm mixed}(p_1,p_2,p_3) ~.
\end{displaymath}

We start computing the diamagnetic contribution $i\gm^{\rm
diam}_{\m_1\m_2\m_3} (p_1,p_2,p_3)$ to the three-point function.  The
one-loop diamagnetic Feynman diagrams that contribute to this function
are shown in fig. 5. The Feynman diagrams in the first row will be
called triangle diagrams, while the remaining diagrams will be
referred to as swordfish diagrams. The sum $i\gm^{\rm
triang}_{\m_1\m_2\m_3} (p_1,p_2,p_3)$ of the triangle diagrams is
formally given in $D$ dimensions by the integral
\begin{equation}
\begin{array}{l}
    {\rm (a) + (b) + (c)} 
        = i\gm^{\rm triang}_{\m_1\m_2\m_3}(p_1,p_2,p_3) \\[6pt]
\hphantom{(a) + (b) + (c)~ }
    {\ds =\,8i\,[D-2] \idqd  \frac{(2q+p_1)_{\m_1} (2q+2p_1+p_2)_{\m_2} 
                            (2q+p_1+p_2)_{\m_3}}
                           {q^2(q+p_1)^2(q+p_1+p_2)^2} } \\[15pt]
\hphantom{(a) + (b) + (c)~}
    {\ds \times\, \sin\!\left[\frac{q\wedge p_1}{2}\right]\,
         \sin\!\left[\frac{(q+p_1)\wedge p_2}{2}\right]\,
         \sin\!\left[\frac{q\wedge (p_1+p_2)}{2}\right]~ .}
\end{array}
\label{triangle}
\end{equation}
The sum $i{\rm I}^{\rm sfish}_{\m_1\m_2\m_3}(p_1,p_2,p_3)$ of the
swordfish diagrams (d) and (g), also in $D$ dimensions, is given by
\begin{equation}
\begin{array}{l}
   {\rm (d) + (g)} =  i{\rm I}^{\rm sfish}_{\m_1\m_2\m_3}
           (p_1,p_2,p_3) \\[6pt] 
\hphantom{(d) + (g)\;}
    {\ds =\, 4i\, [D-2]\,\eta_{\m_1\m_2} \idqd 
         \frac{(2q+p_1+p_2)_{\m_3}}{q^2\,(q+p_1)^2\,(q+p_1+p_2)^2}\,
          \sin\!\left[\frac{q\wedge(p_1+p_2)}{2}\right] }\\[15pt]
\hphantom{(d) + (g)\;} 
   {\ds \times \left\{ \sin\!\left[\frac{(q+p_1)\wedge p_2}{2}\right]\, 
                       \sin\!\left[\frac{q\wedge p_1}{2}\right] 
      +\,\sin\!\left[\frac{(q+p_2)\wedge p_1}{2}\right]\,
         \sin\!\left[\frac{q\wedge p_2}{2}\right]\right\}~, }
\end{array}
\label{swordfpartial}
\end{equation}
and the total swordfish contribution $i\gm^{\rm sfish)}_{\m_1\m_2\m_3}
(p_1,p_2,p_3)$ reads
\begin{displaymath}
\begin{array}{l}
   i\gm^{\rm sfish}_{\m_1\m_2\m_3}(p_1,p_2,p_3)  
      = {\rm (d) + (g) + (e) + (h) + (f) + (i)}\\[6pt]
\hphantom{i\gm^{\rm sfish}_{\m_1\m_2\m_3}(p_1,p_2,p_3)}
   =\, i{\rm I}^{\rm sfish}_{\m_1\m_2\m_3}(p_1,p_2,p_3)
     + i{\rm I}^{\rm sfish}_{\m_3\m_2\m_1}(p_3,p_2,p_1)
     + i{\rm I}^{\rm sfish}_{\m_1\m_3\m_2}(p_1,p_3,p_2)~.
\end{array}
\end{displaymath}
As in the vacuum polarization tensor case, the overall factors
$[\,D-2\,]$ in eqs. (\ref{triangle}) and (\ref{swordfpartial}) account
for the fact that the one-loop diamagnetic contribution is due to the
orbital motion of $D-2$ real scalar quanta in the background field
$B_\m$.  The $D$ comes from the diagrams with fields $Q_\m$
$(\m=1,\ldots,D)$ going around the loop, while the $-2$ is provided by
the corresponding diagrams with ghost loops.  Each diagram in fig. 5
can be expressed as the sum of its planar part and its non-planar part
by repeated use of the identity $2i\sin\!\left(\frac{1}{2}q\wedge
p\right)= e^{iq\wedge p/2}-e^{-iq\wedge p/2}$. The planar part is the
contribution whose $\theta^{\m\n}\!$-dependent complex exponential
factors do not depend on the loop momenta, while the non-planar part
collects all terms whose $\theta^{\m\n}\!$-dependent complex
exponentials depend on the loop momenta. At $D=4$, the planar parts
of the diagrams in fig. 5 are UV divergent by power counting. This UV
divergence is logarithmic and shows in dimensional regularization as a
simple pole at $D=4$. By contrast, the non-planar parts are finite
since the UV divergence is tamed by the the exponentials $e^{iq\wedge
p_i}$, provided $p_i\neq 0$. In other words, the contribution to the
integrals in eqs. (\ref{triangle}) and (\ref{swordfpartial}) of the
quantum modes flowing along the loop with arbitrarily high momenta is
UV divergent when one considers their planar part, and it is turned
into an IR divergence by the non-commutativity of space when one looks
at their non-planar parts.  This IR divergence occurs whenever any of
the conditions
\begin{equation}
   \tp_1=0 \qquad \tp_2=0 \qquad \tp_3=0
\label{3ps}
\end{equation}
is met. This provides a qualitative explanation of the UV/IR for the
diamagnetic part of the three-point function. To make it quantitative,
we need to compute the UV divergent terms in the planar contribution
and the non-commutative IR divergent parts of the non-planar
contribution.

To calculate the UV divergence of the planar contribution, we make a
Laurent expansion about $D=4$ of the corresponding dimensionally
regularized integrals and retain singular contributions. Proceeding in
this way and defining $D=4+2\ee$, we obtain for non-exceptional
external momenta
\begin{displaymath}
\begin{array}{l}
   {\ds i\gm^{\rm trian,P}_{\m_1 \m_2 \m_3}(p_1,p_2,p_3) = 
        \frac{-i}{16\pi^2\ee}\,\frac{2}{3} \,
        2i \sin\!\left(\frac{p_1\wedge p_2}{2}\right) }\\[12pt]
\hphantom{i\gm^{\rm trian,P}_{\m_1 \m_2 m_3}}
   {\ds \times\, \Big[ \eta_{\m_1\m_2}(p_2-p_1)_{\m_3} 
                     + \eta_{\m_1\m_3}(p_1-p_3)_{\m_2} 
                     + \eta_{\m_2\m_3}(p_3-p_2)_{\m_1} \Big]        
                     + O(\ee^0)~, }
\end{array}
\end{displaymath}
for the sum of the triangle diagrams, and
\begin{displaymath}
  i\gm^{\rm sfish,P}_{\m_1 \m_2 \m_3}(p_1,p_2,p_3)=O(\ee^0) 
\end{displaymath}
for the sum of the swordfish diagrams. From this we conclude that the
UV divergent contribution $i\gm^{\rm
diam,UV}_{\m_1\m_2\m_3}(p_1,p_2,p_3)$ of the diamagnetic part to the
three-point function is given in dimensional regularization by
\begin{equation}
\begin{array}{l}
   {\ds i\gm^{\rm diam,UV}_{\m_1 \m_2 \m_3}(p_1,p_2,p_3) = 
        \frac{-i\,a^{\rm diam}_1}{\ee}\,2i
        \sin\!\left(\frac{p_1\wedge p_2}{2}\right) }\\[12pt]
\hphantom{ i\gm^{\rm diam,UV}_{\m_1 \m_2 \m_3}}
   {\ds \times\,\Big[\, \eta_{\m_1\m_2}(p_2-p_1)_{\m_3} 
                     + \eta_{\m_1\m_3}(p_1-p_3)_{\m_2} 
                     + \eta_{\m_2\m_3}(p_3-p_2)_{\m_1} \Big] ~, }
\end{array}
\label{ThreediamUV}
\end{equation}
where $a^{\rm diam}_1$ is as in eq. (\ref{coeffnon}). To compute the IR
divergences that appear in the non-planar contribution when one or
more of the $\tp_i$ vanish, we set $D=4$ and use the following
basic results for $\tk\to 0$:
\begin{eqnarray}
  &{\ds \idq \frac{e^{i\,q\wedge k}}{q^2\,(q+\ell_1)^2} \approx
       \frac{-i}{16\pi^2}\; \ln\tilde{k}^2 } & 
    \label{IR3pintegrala} \\[12pt]
  &{\ds \idq \frac{q_{\m_1}q_{\m_2}\,e^{i\,q\wedge k}}
                 {q^2\,(q+\ell_1)^2\,(q+\ell_2)^2} 
       \approx \frac{-i}{16\pi^2}\; 
       \frac{1}{4}\, \ln\tilde{k}^2\, \eta_{\m_1\m_2} }& 
       \label{IR3pintegralb}
\end{eqnarray}
\begin{equation}
\begin{array}{l}
   {\ds \idq \frac{q_{\m_1}q_{\m_2}q_{\m_3}~e^{i\,q\wedge k}}
                  {q^2\,(q+\ell_1)^2\,(q+\ell_2)^2} \approx 
        \frac{i}{16\pi^2} }\\[12pt]
\hphantom{\hspace{75pt}}
   {\ds \times\,\bigg\{ \frac{i}{2\,\tk^2} \left[
        \eta_{\m_1\m_2}\,\tk_{\m_3} + \eta_{\m_2\m_3}\,\tk_{\m_1}
        + \eta_{\m_1\m_3}\,\tk_{\m_2}
        - 2\,\frac{\,\tk_{\m_1}\tk_{\m_2}\tk_{\m_3}}{\tk^2}\right] 
   }\\[12pt]
\hphantom{\hspace{75pt} }
  {\ds  +\, \frac{1}{12}\,\ln\tk^2\, 
          \Big[\, \eta_{\m_1\m_2}\,(\ell_1+\ell_2)_{\m_3}
               + \eta_{\m_2\m_3}\,(\ell_1+\ell_2)_{\m_1}
               + \eta_{\m_1\m_3}(\ell_1+\ell_2)_{\m_2}\,\Big]
          \bigg\} ~,}
\end{array}
\label{IR3pintegralc}
\end{equation}
where $\ell_1$ and $\ell_2$ never vanish and $\approx$ indicates that we
have dropped any contribution which remains finite as $\tk\to
0$. These expressions, together with some algebra, lead to the
following divergences for the non-planar parts of the diagrams in
fig. 5 when one or more $\tp_i$ approach zero:
\begin{displaymath}
\begin{array}{l}
   {\ds i\gm_{\m_1\m_2\m_3}^{\rm triang,NP}(p_1,p_2,p_3) 
        \approx \frac{i}{16\pi^2}\,\frac{2}{9}
        \,2i\sin\!\left(\frac{p_1\wedge p_2}{2}\right)\,
        \left( \ln \tp_1^2 + \ln \tp_2^2 
             + \ln \tp_3^2 \right) }\\[12pt] 
\hphantom{i\gm_{\m_1}}  
  {\ds \times \,\Big[ \eta_{\m_1\m_2}(p_2-p_1)_{\mu_3} 
             + \eta_{\m_1\m_3}(p_1-p_3)_{\mu_2}
             + \eta_{\m_2\m_3}(p_3-p_2)_{\mu_1} \Big] }\\[9pt]
\hphantom{i\gm_{\m_1}}  
   {\ds +\,\frac{1}{\pi^2}\,
        \cos\left(\frac{p_1\wedge p_2}{2}\right) \sum_{i=1}^3\,
        \left[\, \frac{2\,(\tp_i)_{\m_1}(\tp_i)_{\m_2}(\tp_i)_{\m_3}}   
                    {(\tp^2_i)^2} 
             - \eta_{\m_1\m_2}\frac{(\tp_i)_{\m_3}}{\tp^2_i} 
             - \eta_{\m_2\m_3}\frac{(\tp_i)_{\m_1}}{\tp^2_i}
             - \eta_{\m_1\m_3}\frac{(\tp_i)_{\m_2}}{\tp^2_i}\,\right]
    }
\end{array}
\end{displaymath}
and 
\begin{displaymath}
  i\gm_{\m_1\m_2\m_3}^{\rm sfish,NP}(p_1,p_2,p_3) \approx 
       \frac{1}{\pi^2}\,\cos\left(\frac{p_1\wedge p_2}{2}\right)
       \sum_{i=1}^3\,\left[  
          \eta_{\m_1\m_2}\frac{(\tp_i)_{\m_3}}{\tp^2_i} 
        + \eta_{\m_2\m_3}\frac{(\tp_i)_{\m_1}}{\tp^2_i}
        + \eta_{\m_1\m_3}\frac{(\tp_i)_{\m_2}}{\tp^2_i}\,\right]~.
\end{displaymath}
Adding these two expressions we obtain for the non-commutative
IR divergent part of the diamagnetic contribution to the three-point
function
\begin{equation}
\begin{array}{l}
   {\ds i\gm_{\m_1\m_2\m_3}^{\rm diam,IR}(p_1,p_2,p_3) = 
        \frac{2}{\pi^2}\,
        \cos\left(\frac{p_1\wedge p_2}{2}\right) \sum_{i=1}^3\,
        \frac{(\tp_i)_{\m_1}(\tp_i)_{\m_2}(\tp_i)_{\m_3}}   
             {(\tp^2_i)^2} }\\[12pt]
\hphantom{i\gm_{\m_1\m_2\m_3}^{\rm diam,IR}(p_1,p_2,p_3)~}
   {\ds +\; \frac{i}{3}\, a^{\rm diam}\,2i 
        \sin\!\left(\frac{p_1\wedge p_2}{2}\right)\,
        \left( \ln \tp_1^2 + \ln \tp_2^2 
                          + \ln \tp_3^2 \right) }\\[12pt] 
\hphantom{i\gm_{\m_1\m_2\m_3}^{\rm diam,IR}(p_1,p_2,p_3)~}  
  {\ds \times \,\Big[ \eta_{\m_1\m_2}(p_2-p_1)_{\mu_3} 
             + \eta_{\m_1\m_3}(p_1-p_3)_{\mu_2}
             + \eta_{\m_2\m_3}(p_3-p_2)_{\mu_1} \Big]~, }
\end{array}
\label{threediamIR}
\end{equation}
where $ a^{\rm diam}_1$ is given in eq. (\ref{coeffnon}). Notice that
the logarithmic term in this expression is dual to the the UV
divergent contribution given in eq. (\ref{ThreediamUV}), in the sense
that in the momenta region
\begin{equation}
   |\tp_1| \sim |\tp_2| \sim |\tp_3| \sim \theta\Lambda_{\rm IR} \to 0
\label{3plog}
\end{equation}
the former can be obtained from the latter by using the definitions
(\ref{identifications}).  Note also that the factor
$\,\sin\!\left(\frac{1}{2}\,(p_1\wedge p_2\right)$, together with
momentum conservation $\tp_1+\tp_2+\tp_3=0$, renders the second term
in eq. (\ref{threediamIR}) finite when $\tp_i\to 0$ $(i=1,2,3)$. This
is so in spite of the logarithmic non-commutative IR divergences of
the integrals (\ref{IR3pintegrala})-(\ref{IR3pintegralc}) that enter
the computation of the non-planar part of the diamagnetic
contribution, and can be thought of as a consequence of gauge
invariance --which determines the momentum structure of the UV
divergent part of the three-point function in
eq. (\ref{ThreediamUV})-- and the UV/IR mixing mechanism. Indeed,
UV/IR mixing in the form of eq. (\ref{identifications}) allows to
retrieve from the UV divergent contribution $i\gm_{\m_1\m_2\m_3}^{\rm
diam,UV}(p_1,p_2,p_3)$ the terms in $i\gm_{\m_1\m_2\m_3}^{\rm
diam,NP}(p_1,p_2,p_3)$ which involve $\ln\tp_i^2$, with $\tp_i$
verifying eq. (\ref{3plog}).

Let us compute now the paramagnetic part $i\gm^{\rm
param}_{\m_1\m_2\m_3}(p_1,p_2,p_3)$ of the the three-point
function. This is given by the contribution involving only three
background fields that comes from the diagram in fig. 4, which in turn
is given by eqs. (\ref{FFparam}) and (\ref{Jofp}). Extracting from
eq. (\ref{FFparam}) the contribution cubic in the background fields
and using the results of section 2 for the integral $J(p)$, we obtain,
after some calculations,
\begin{equation}
\begin{array}{l}
   {\ds i\gm^{\rm param,UV}_{\m_1\m_2\m_3}(p_1,p_2,p_3) = 
        \frac{-i\, a^{\rm param}}{\ee}\,
        2i\,\sin\!\left(\frac{p_1\wedge p_2}{2}\right) }\\[12pt]
\hphantom{i\gm^{\rm param,UV}_{\m_1\m_2\m_3}(p_1,p_2,p_3)~}
  {\ds \times \,\Big[ \eta_{\m_1\m_2}(p_2-p_1)_{\mu_3} 
             + \eta_{\m_1\m_3}(p_1-p_3)_{\mu_2}
             + \eta_{\m_2\m_3}(p_3-p_2)_{\mu_1} \Big] }
\end{array}
\label{ThreeparUV}
\end{equation} 
for the UV divergent part, and
\begin{equation}
\begin{array}{l}
   {\ds i\gm^{\rm param,IR}_{\m_1\m_2\m_3}(p_1,p_2,p_3) = 
      \frac{i}{3}\; a^{\rm param}\;2i\,
      \sin\!\left(\frac{p_1\wedge p_2}{2}\right)\,  
      \left( \ln \tp_1^2 + \ln \tp_2^2 
                          + \ln \tp_3^2 \right)}\\[6pt]
\hphantom{i\gm^{\rm param,IR}_{\m_1\m_2\m_3}(p_1,p_2,p_3)~}
   {\ds \times \,\Big[ \eta_{\m_1\m_2}(p_2-p_1)_{\mu_3} 
              + \eta_{\m_1\m_3}(p_1-p_3)_{\mu_2}
              + \eta_{\m_2\m_3}(p_3-p_2)_{\mu_1} \Big] }
\end{array}
\label{threeparamIR}
\end{equation}
for the IR divergences that appear as $\tp_i\to 0$. Here $a^{\rm
param}_1$ is defined in eq. (\ref{coeffnon}). As in the case of the
diamagnetic contribution, the IR part $i\gm^{\rm
param,IR}_{\m_1\m_2\m_3}(p_1,p_2,p_3)$ is dual to the UV part
$i\gm^{\rm param,UV}_{\m_1\m_2\m_3}(p_1,p_2,p_3)$, in the sense that
one can go from one to another by using the identifications
(\ref{identifications}) in the momentum region (\ref{3plog}).

We finally look at the mixed part $i\gm^{\rm
mixed}_{\m_1\m_2\m_3}(p_1,p_2,p_3)$. It receives contributions from
the diagrams in figs. 2 and 6. As explained in section 2, the diagram
of fig. 2 vanishes. The very same arguments as for the diagram in
fig. 2 show that the diagram in fig. 6(a) also vanishes. By contrast,
the diagrams in figs, 6(b) and 6(c) are different from zero; yet their
planar parts are finite by power counting and their non-planar parts
are IR finite for vanishing $\tp_i$. Altogether, we conclude
\begin{equation}
   \gm^{\rm mixed,UV}_{\m_1\m_2\m_3}(p_1,p_2,p_3) 
     =  \gm^{\rm mixed,IR}_{\m_1\m_2\m_3}(p_1,p_2,p_3) = 0~.
\label{3mixed}
\end{equation}

The UV divergent part $i\gm^{\rm UV}_{\m_1\m_2\m_3}(p_1,p_2,p_3)$ of
the three-point function will be the sum of the diamagnetic,
paramagnetic and mixed UV divergent parts. Using the results obtained,
we have 
\begin{equation}
\begin{array}{l}
   {\ds i\gm^{\rm UV}_{\m_1\m_2\m_3}(p_1,p_2,p_3) = \frac{-i}{\ee}\, 
        \left(a^{\rm diam} + a^{\rm param} \right) \,
        2i \sin\!\left(\frac{p_1\wedge p_2}{2}\right) }\\[12pt]
\hphantom{i\gm^{\rm UV}_{\m_1\m_2\m_3}(p_1,p_2,p_3)~}
   {\ds \times \,\Big[ \eta_{\m_1\m_2}(p_2-p_1)_{\mu_3} 
              + \eta_{\m_1\m_3}(p_1-p_3)_{\mu_2}
              + \eta_{\m_2\m_3}(p_3-p_2)_{\mu_1} \Big] ~,}
\end{array}
\label{threepUV}
\end{equation} 
with $a^{\rm diam}$ and $a^{\rm param}$ as in eq. (\ref{coeffnon}).
It is clear that the counterterm in eq. (\ref{counterterm}) subtracts
the UV divergences in the three-point function and that the beta
function computed from the three-point function is also given by
eq. (\ref{nonbeta}), all this in accordance with gauge invariance. We
again see that the negative sign of the one-loop beta function is due
to the dominance of the paramagnetic high-momentum contributions over
the diamagnetic ones. We close this section displaying the leading
non-commutative IR behaviour of the three-point function, obtained by
summing the contributions in eqs. (\ref{threediamIR}) and
(\ref{threeparamIR}):
\begin{equation}
\begin{array}{l}
   {\ds i\gm_{\m_1\m_2\m_3}^{\rm IR}(p_1,p_2,p_3) = 
        \frac{2}{\pi^2}\,
        \cos\left(\frac{p_1\wedge p_2}{2}\right) \sum_{i=1}^3\,
        \frac{(\tp_i)_{\m_1}(\tp_i)_{\m_2}(\tp_i)_{\m_3}}   
             {(\tp^2_i)^2} }\\[12pt] 
\hphantom{i\gm_{\m_1\m_2\m_3}^{\rm IR}(p_1,p_2,p_3)~}  
   {\ds + \;\frac{i}{3} \left(a^{\rm diam} + a^{\rm param}\right) 
        \,2i \sin\!\left(\frac{p_1\wedge p_2}{2}\right)\,
        \left( \ln \tp_1^2 + \ln \tp_2^2 
                          + \ln \tp_3^2 \right) }\\[12pt] 
\hphantom{i\gm_{\m_1\m_2\m_3}^{\rm IR}(p_1,p_2,p_3)~}  
  {\ds \times \,\Big[ \eta_{\m_1\m_2}(p_2-p_1)_{\mu_3} 
             + \eta_{\m_1\m_3}(p_1-p_3)_{\mu_2}
             + \eta_{\m_2\m_3}(p_3-p_2)_{\mu_1} \Big]. }
\end{array}
\label{threepIR}
\end{equation}
Notice that the non-logarithmic non-commutative IR contributions to
the three-point are only supplied by its diamagnetic part [see
eqs. (\ref{threediamIR}) and (\ref{threeparamIR})]; an effect that was
met for the first time when we studied the vacuum polarization tensor.

\section{The four-point function}

In this section we compute the UV divergent and the leading
non-commutative IR part of the four-point function and exhibit how
UV/IR connects them. The four-point function $i\gm_{\m_1\m_2\m_3\m_4}
(x_1,x_2,x_3,x_4)$ is defined by
\begin{equation}
   i\gm_{\m_1\m_2\m_3\m_4}(x_1,x_2,x_3,x_4) = \frac{\d^4 \Gamma[B]}
   {\d B_{\m_1}(x_1)\d B_{\m_2}(x_2) 
   \d B_{\m_3}(x_3)\d B_{\m_4}(x_4)}{\bigg\arrowvert}_{B=0}~,
\end{equation}        
where $i\Gamma[B]=\ln Z[B]$ and $Z[B]$ is as in
eq. (\ref{uonepartition}). In momentum space we will write
$i\gm_{\m_1\m_2\m_3\m_4}(p_1,p_2,p_3,p_4)$, with $p_1+p_2+p_3+p_4=0$,
and will be the sum of a diamagnetic, a paramagnetic and a mixed
contribution:
\begin{displaymath}
\begin{array}{l}
   i\gm_{\m_1\m_2\m_3\m_4}(p_1,p_2,p_3,p_4) 
      = i\gm_{\m_1\m_2\m_3\m_4}^{\rm diam}(p_1,p_2,p_3,p_4) \\[6pt]
\hphantom{ i\gm_{\m_1\m_2\m_3\m_4}(p_1,p_2,p_3,p_4) \,}
      +\, i\gm_{\m_1\m_2\m_3\m_4}^{\rm param}(p_1,p_2,p_3,p_4)     
      + i\gm_{\m_1\m_2\m_3\m_4}^{\rm mixed}(p_1,p_2,p_3,p_4) ~.
\end{array}
\end{displaymath}
Let us consider each one of these contributions separately.

The diamagnetic contribution is given by the sum of the diagrams in
fig. 7. The diagrams in the first, second and third row of the figure
have different topologies and will be called box, lynx and bubble
diagrams respectively. Each one of these diagrams is a sum of a planar
part and a non-planar part. The same arguments as in sections 2 and 3
show that the planar parts are UV divergent by power counting at
$D=4$, the divergence being logarithmic and being characterized as a
pole in dimensional regularization.  As regards the non-planar parts,
these are finite for $\tp_i\neq 0$ and develop divergences for any of
the configurations
\begin{equation}
    \tp_1=0 \quad \tp_2=0 \quad \tp_3=0 \quad \tp_4=0 \quad
    \tp_1+\tp_2=0 \quad \tp_1+\tp_3=0 \quad \tp_2+\tp_3=0 ~. 
\label{4pIRregion}
\end{equation}
To calculate the UV divergences, we make a Laurent expansion about
$D=4$ of the dimensionally regularized integrals defining the planar
parts and obtain, for non-exceptional momenta:
\begin{equation}
\begin{array}{l}
   {\ds i\gm^{\rm box,P}_{\m_1\m_2\m_3\m_4}(p_1,p_2,p_3,p_4) = 
        \frac{-i}{16\pi^2\ee} ~ \frac{8}{3} ~
         \big( \eta_{\m_1\m_2} \eta_{\m_3\m_4}
                       + \eta_{\m_1\m_3} \eta_{\m_2\m_4}
                       + \eta_{\m_1\m_4} \eta_{\m_2\m_3} 
                  \big) }\\[12pt]
\hphantom{i\gm^{\rm box,P}_{\m_1\m_2\m_3\m_4}~} 
  {\ds \times \, \bigg[ \cosppp + \cosppm }\\[12pt]
\hphantom{i\gm^{\rm box,P}_{\m_1\m_2\m_3\m_4} \times \,}    
   {\ds +\, \cospmm \bigg] + O(\ee^0) }
\end{array}
\label{boxUV}
\end{equation}
for the sums of the of the box diagrams, and
\begin{equation}
\begin{array}{l}
   {\ds i\gm^{\rm lynx,P}_{\m_1\m_2\m_3\m_4}(p_1,p_2,p_3,p_4)\,=\,
        -\;\frac{1}{2}~
        i\gm^{\rm bub,P}_{\m_1\m_2\m_3\m_4}(p_1,p_2,p_3,p_4) 
        + \,O(\ee^0) }\\[12pt]
\hphantom{i\gm^{\rm lynx,P}_{\m_1\m_2\m_3\m_4} }
   {\ds = \frac{i}{16\pi^2\ee} \; 8 \,\bigg[ 
         \big( \eta_{\m_1\m_2} \eta_{\m_3\m_4} 
         + \eta_{\m_1\m_4} \eta_{\m_2\m_3} \big)\cosppp  }\\[12pt]
\hphantom{i\gm^{\rm lynx,P}_{\m_1\m_2\m_3\m_4} 
          = \frac{i}{16\pi^2\ee} \; 8 \, }
   {\ds +\, \big( \eta_{\m_1\m_3} \eta_{\m_2\m_4} 
            + \eta_{\m_1\m_4} \eta_{\m_2\m_3} \big) \cosppm }\\[12pt] 
\hphantom{i\gm^{\rm lynx,P}_{\m_1\m_2\m_3\m_4} 
          = \frac{i}{16\pi^2\ee} \; 8 \, }
   {\ds +\, \big( \eta_{\m_1\m_2} \eta_{\m_3\m_4} 
                 + \eta_{\m_1\m_3} \eta_{\m_2\m_4} 
            \big) \cospmm \bigg] + O(\ee^0) ~.}
\end{array}
\label{lynxUV}
\end{equation}
for the sum of the lynx and bubble diagrams.  Here, as in previous
sections, $D=4+2\ee$. Summing these three contributions and doing some
simple trigonometry, we arrive at the following UV divergences
for the diamagnetic part of the four-point function:
\begin{equation}
\begin{array}{l}
  {\ds i\gm^{\rm diam,UV}_{\m_1\m_2\m_3\m_4}(p_1,p_2,p_3,p_4) = 
       \frac{i\, a^{\rm diam}}{\ee} \,
       \bigg[ \big( \etaonethreetwofour-\etaonefourtwothree \big) 
              \sin\!\left( \frac{p_1\wedge p_2}{2}\right) 
              \sin\!\left( \frac{p_3\wedge p_4}{2}\right)}\\[12pt]
\phantom{i\gm^{\rm diam,UV}_{\m_1\m_2\m_3\m_4}(p_1,p_2,p_3,p_4) = 
         \frac{i\, a^{\rm diam}}{\ee} ~}
  {\ds +\, \big( \etaonethreetwofour-\etaonetwothreefour \big)
           \sin\!\left( \frac{p_1\wedge p_4}{2}\right) 
           \sin\!\left( \frac{p_3\wedge p_2}{2}\right) }\\[12pt]
\phantom{i\gm^{\rm diam,UV}_{\m_1\m_2\m_3\m_4}(p_1,p_2,p_3,p_4) =
         \frac{i\, a^{\rm diam}}{\ee} ~}
  {\ds +\, \big( \etaonethreetwofour-\etaonetwothreefour \big)
           \sin\!\left( \frac{p_4\wedge p_2}{2}\right) 
           \sin\!\left( \frac{p_3\wedge p_1}{2}\right) \bigg]~,}
\end{array}
\label{4pointdiamUV}
\end{equation}
with $a^{\rm diam}$ given in eq. (\ref{coeffnon}). Let us stress that,
unlike for the three-point function, the trigonometric and tensor
structure of the box, lynx and bubble contributions is different from
that in the classical action $S_{\rm class}[B]$ for the quartic vertex
$B^4$, and that it is precisely the sum over diagrams with different
topologies what produces the correct structure, thus making
non-trivial the answer given by eq. (\ref{4pointdiamUV}). Of course,
this is so because gauge invariance is at work and brings the
necessary simplifications.

We next compute the leading non-commutative IR behaviour of the
non-planar parts of the diagrams in fig. 7. To do this, we use
eqs. (\ref{IR3pintegrala}) and (\ref{IR3pintegralb}) and the result
\begin{displaymath}
\begin{array}{l}
  {\ds \idq \frac{q_{\m_1}q_{\m_2}q_{\m_3}q_{\m_4}~e^{i\,q\wedge k}}
                  {q^2\,(q+\ell_1)^2\,(q+\ell_2)^2\,(q+\ell_3)^2} 
        \approx }\\[12pt]
\hphantom{ \idq \frac{1}{q^2\,(q+\ell_1)^2}~~}
  {\ds \approx \frac{-i}{16\pi^2}\;\frac{1}{24}\,\ln\tk^2 
       \left( \etaonetwothreefour + \etaonethreetwofour 
             + \etaonefourtwothree \right) }
\end{array}
\end{displaymath}
for $\tk\to 0$, where $\ell_1,~\ell_2$ and $\ell_3$ never vanish and
$\approx$ indicates that we have dropped contributions which remain
finite as $\tk\to 0$. After some algebra, we obtain for the three sets
of diagrams under consideration:
\begin{equation}
\begin{array}{l}
   {\ds i\gm^{\rm box,NP}_{\m_1\m_2\m_3\m_4}(p_1,p_2,p_3,p_4) \approx 
        \frac{-i}{16\pi^2} \, \big( \etaonetwothreefour 
        + \etaonethreetwofour + \etaonefourtwothree \big) }\\[12pt]
\hphantom{i\gm^{\rm box)}}
   {\ds \times \, 4\, \bigg\{ \cosppp \bigg[ \ln\vs + \ln\vt 
                       - \frac{2}{3}\, \sumlogs\, \bigg] }\\[12pt]
\hphantom{i\gm^{\rm box)}\times ~}
   {\ds +\, \cosppm \bigg[ \ln\vt + \ln\vu 
                       - \frac{2}{3}\, \sumlogs \,\bigg] }\\[12pt]
\hphantom{i\gm^{\rm box)}\times ~}
   {\ds +\, \cosppm \bigg[ \ln\vs + \ln\vu
                       - \frac{2}{3}\, \sumlogs \,\bigg] \bigg\} }
\end{array}
\label{boxIR}
\end{equation}
and
\begin{equation}
\begin{array}{l}
  {\ds i\gm^{\rm lynx,NP}_{\m_1\m_2\m_3\m_4}(p_1,p_2,p_3,p_4) 
       \,\approx\, -\,\frac{1}{2}~
        i\gm^{\rm bub,NP}_{\m_1\m_2\m_3\m_4}(p_1,p_2,p_3,p_4) }\\[12pt]
\hphantom{i\gm_{\m}}
   {\ds  \approx \frac{-i}{16\pi^2}\; 8 \, \bigg\{ \bigg[ \big( 
        \etaonetwothreefour + \etaonefourtwothree \big)\, \cosppp }\\[12pt]
\hphantom{i\gm^{\rm lynx,NP}_{\m\m_2\m_3\m_4}\times 8\;}
   {\ds +\,\big( \etaonetwothreefour + \etaonethreetwofour \big)\, 
        \cospmm }\\[12pt]
\hphantom{i\gm^{\rm lynx,NP}_{\m\m_2\m_3\m_4}\times 8\;}
   {\ds +\,\big( \etaonethreetwofour + \etaonefourtwothree \big)\,
        \cosppm \!\bigg] \,\sum_{i=1}^4 \,\ln\tp_i^2 }\\[12pt]
\hphantom{i\gm^{\rm lynx,NP}_{\m\m_2\m_3\m_4} 8~\; }
   {\ds -\;\bigg[\, \big( \etaonetwothreefour 
        + 2\,\etaonefourtwothree \big) \,\cosppp }\\[12pt] 
\hphantom{i\gm^{\rm lynx,NP}_{\m\m_2\m_3\m_4} \times 8\; }
   {\ds +\,\big( \etaonetwothreefour + 2\,\etaonethreetwofour \big) \,
        \cospmm \!\bigg] \ln\vs }\\[12pt]
\hphantom{i\gm^{\rm lynx,NP}_{\m\m_2\m_3\m_4} 8~\; }
   {\ds -\;\bigg[\,  \big( 2\,\etaonetwothreefour 
        + \etaonefourtwothree \big)\, \cosppp }\\[12pt]
\hphantom{i\gm^{\rm lynx,NP}_{\m\m_2\m_3\m_4} \times 8\; }
   {\ds +\, \big( 2\etaonetwothreefour + \etaonefourtwothree \big)\,
        \cosppm \!\bigg] \ln\vt }\\[12pt]
\hphantom{i\gm^{\rm lynx,NP}_{\m\m_2\m_3\m_4} 8 ~\; }
   {\ds -\;\bigg[\,  \big( 2\,\etaonetwothreefour 
        + \etaonethreetwofour \big) \,\cospmm }\\[12pt]
\hphantom{i\gm^{\rm lynx,NP}_{\m\m_2\m_3\m_4} \times 8\; }
   {\ds +\, \big( \etaonethreetwofour + 2\,\etaonefourtwothree \big)\,
        \cospmm\! \bigg] \ln\vu \bigg\} \;.}
\end{array}
\label{lynxIR}
\end{equation}
Summing these three results we obtain for the leading non-commutative
IR diamagnetic contribution to the four-point function
\begin{equation}
\begin{array}{l}
  {\ds i\gm^{\rm diam,IR}_{\m_1\m_2\m_3\m_4}(p_1,p_2,p_3,p_4) = 
       - 4i\, a^{\rm diam}}\\[6pt]
\hphantom{i\gm_{\m_1}}
  {\ds \times\, \bigg\{ \big( \etaonethreetwofour 
                            - \etaonefourtwothree \big) 
                \sin\!\left( \frac{p_1\wedge p_2}{2}\right) 
                \sin\!\left( \frac{p_3\wedge p_4}{2}\right) 
                \bigg[ \sumlogs - 3\ln\vs \bigg] }\\[12pt]
\hphantom{i\gm_{\m_1} \times}
  {\ds +\, \big( \etaonethreetwofour-\etaonetwothreefour \big)
           \sin\!\left( \frac{p_1\wedge p_4}{2}\right) 
           \sin\!\left( \frac{p_3\wedge p_2}{2}\right) 
           \bigg[ \sumlogs - 3\ln\vt \bigg] }\\[12pt]
\hphantom{i\gm_{\m_1} \times}
  {\ds +\, \big( \etaonethreetwofour-\etaonetwothreefour \big)
           \sin\!\left( \frac{p_4\wedge p_2}{2}\right) 
           \sin\!\left( \frac{p_3\wedge p_1}{2}\right) 
           \bigg[ \sumlogs - 3\ln\vu \bigg] \bigg\} \,,}\\[12pt]
\end{array}
\label{4pointdiamIR}
\end{equation}
$a^{\rm diam}$ being given in eq. (\ref{coeffnon}). Two comments are
now in order. First, eqs. (\ref{boxIR}) and (\ref{lynxIR}) show that
the non-planar parts of the four-point diamagnetic diagrams exhibit IR
divergences for configurations (\ref{4pIRregion}). Yet, the full
diamagnetic contribution, given by eq. (\ref{4pointdiamIR}), is free
of such singularities. There is, therefore, a cancellation mechanism at
work among the non-commutative IR divergent contributions coming from
the box, lynx and bubble diagrams. This mechanism is a consequence of
gauge invariance being preserved at one-loop. Let us substantiate this
statement. As we shall see in the next section, gauge invariance leads
to the following Ward identity involving the three and four-point
functions:
\begin{equation}
\begin{array}{l}
   {\ds p_4^{\m_4}\,\gm_{\m_1\m_2\m_3\m}(p_1,p_2,p_3,p_4) =
        2 \sin\!\left(\frac{p_1\wedge p_4}{2}\right) 
        \gm_{\m_1\m_2\m_3}(-p_2\!-\!p_3,p_2,p_3)}\\[12pt]
\hphantom{p_4^{\m_4}\,\gm_{\m_1\m_2\m_3\m}(p_1,p_2,p_3,p_4)~\,}
   {\ds +\, 2 \sin\!\left(\frac{p_2\wedge p_4}{2}\right) 
          \gm_{\m_1\m_2\m_3}(-p_1\!-\!p_3,p_3,p_1)}\\[12pt]
\hphantom{p_4^{\m_4}\,\gm_{\m_1\m_2\m_3\m}(p_1,p_2,p_3,p_4)~\,}
   {\ds +\, 2 \sin\!\left(\frac{p_3\wedge p_4}{2}\right) 
          \gm_{\m_1\m_2\m_3}(-p_1\!-\!p_2,p_1,p_2) ~. }
\end{array}
\label{fourthreegaugeeq}
\end{equation}
Now, the r.h.s of this equation is finite for any of the
configurations (\ref{4pIRregion}), as can be readily shown by
substituting eq. (\ref{threepIR}) in it. Thus, if
$\Gamma_{\m_1\m_2\m_3\m_4} (p_1,p_2,p_3,p_4)$ had an IR
diamagnetic divergence for momenta (\ref{4pIRregion}), it would have
to be transverse with respect to $p_4^{\m_4}$. However, there is no
tensor transverse to $p_4^{\m_4}$ that can be built with the tensors
and functions that occur in eqs. (\ref{boxIR}) and (\ref{lynxIR}).
Secondly, the contribution to $i\gm^{\rm diam,NP}_{\m_1\m_2\m_3\m_4}
(p_1,p_2,p_3,p_4)$ coming from the momentum region
\begin{equation}
   |\tp_1|\sim |\tp_2| \sim |\tp_3| \sim |\tp_4| \sim
   |\tp_1+\tp_2|\sim |\tp_1+\tp_3| \sim |\tp_2+\tp_3| \sim 
   \theta\Lambda_{\rm IR} \to 0
\label{4plog}
\end{equation}
is           
\begin{equation}
\begin{array}{l}
  {\ds i\gm^{\rm diam,NP}_{\m_1\m_2\m_3\m_4}(p_1,p_2,p_3,p_4) 
       \approx - 4i\,a^{\rm diam}\,\ln(\th\Lambda_{\rm IR})^2 }\\[6pt]
\hphantom{i\gm^{\rm diam,NP}_{\m_1\m_2\m_3\m_4}~}
  {\ds \times\, \bigg[ \big( \etaonethreetwofour 
                            - \etaonefourtwothree \big) 
                \sin\!\left( \frac{p_1\wedge p_2}{2}\right) 
                \sin\!\left( \frac{p_3\wedge p_4}{2}\right) }\\[12pt]
\hphantom{i\gm^{\rm diam,NP}_{\m_1\m_2\m_3\m_4} \times\,}
  {\ds +\, \big( \etaonethreetwofour-\etaonetwothreefour \big)
           \sin\!\left( \frac{p_1\wedge p_4}{2}\right) 
           \sin\!\left( \frac{p_3\wedge p_2}{2}\right) }\\[12pt]
\hphantom{i\gm^{\rm diam,NP}_{\m_1\m_2\m_3\m_4} \times\,}
  {\ds +\, \big( \etaonethreetwofour-\etaonetwothreefour \big)
           \sin\!\left( \frac{p_4\wedge p_2}{2}\right) 
           \sin\!\left( \frac{p_3\wedge p_1}{2}\right) \bigg] ~.}
\end{array}
\label{IRlogarithm}
\end{equation}
As happened for the two-point and three-point functions, this
logarithmic IR contribution can be retrieved from the UV divergent
contribution $i\gm^{\rm diam,UV}_{\m_1\m_2\m_3\m_4} (p_1,p_2,p_3,p_4)$
by performing the identifications (\ref{identifications}). This
provides an explicit realization for the four-point function of the
UV/IR mixing characteristic of quantum field theories on
non-commutative Minkowski space-time. This UV/IR mixing works for each
single Feynman diagram: note that one can apply the identifications
(\ref{identifications}) to eqs. (\ref{boxUV}) and (\ref{lynxUV}) to
obtain eqs. (\ref{boxIR}) and (\ref{lynxIR}) for momenta verifying the
conditions (\ref{4plog}).

We now come to the paramagnetic part of the four-point function. It is
given by the contribution quartic in the background field that comes
from the diagram in fig. 4, the latter being given by
eqs. (\ref{FFparam}) and (\ref{Jofp}). As already explained, the
planar part of the diagram con\-tains all UV divergences, while the
non-planar part collects all integrals with non-commutative IR
singularities. Projecting out from eq. (\ref{FFparam}) the
contribution quartic in $B_\m$ and using the results of section 2 for
$J(p)$, we obtain the following UV divergences and leading
non-commutative IR terms in the paramagnetic contribution to the
four-point function:
\begin{equation}
\begin{array}{l}
  {\ds i\gm^{\rm param,UV}_{\m_1\m_2\m_3\m_4}(p_1,p_2,p_3,p_4) = 
       \frac{i\, a^{\rm param}}{\ee}}\\[9pt]
\hphantom{i\gm^{\rm param,UV}_{\m_1\m_2\m_3\m_4} }
  {\ds \times\, \bigg[\, 
       \big( \etaonethreetwofour-\etaonefourtwothree \big) 
       \sin\!\left( \frac{p_1\wedge p_2}{2}\right) 
       \sin\!\left( \frac{p_3\wedge p_4}{2}\right)}\\[12pt]
\hphantom{i\gm^{\rm param,UV}_{\m_1\m_2\m_3\m_4}\times}
  {\ds +\, \big( \etaonethreetwofour-\etaonetwothreefour \big)
           \sin\!\left( \frac{p_1\wedge p_4}{2}\right) 
           \sin\!\left( \frac{p_3\wedge p_2}{2}\right) }\\[12pt]
\hphantom{i\gm^{\rm param,UV}_{\m_1\m_2\m_3\m_4}\times}
  {\ds +\, \big( \etaonethreetwofour-\etaonetwothreefour \big)
           \sin\!\left( \frac{p_4\wedge p_2}{2}\right) 
           \sin\!\left( \frac{p_3\wedge p_1}{2}\right) \bigg] }
\end{array}
\label{4pointparamUV}
\end{equation}
and
\begin{equation}
\begin{array}{l}
  {\ds i\gm^{\rm param,IR}_{\m_1\m_2\m_3\m_4}(p_1,p_2,p_3,p_4) = 
       -\,4i\,a^{\rm param} }\\[9pt]
\hphantom{i\gm^{\rm param,IR}_{\m_1\m_2\m_3\m_4} }
  {\ds \times\, \bigg[\, 
       \big( \etaonethreetwofour-\etaonefourtwothree \big) 
       \sin\!\left( \frac{p_1\wedge p_2}{2}\right) 
       \sin\!\left( \frac{p_3\wedge p_4}{2}\right)
       \ln\vs}\\[12pt]
\hphantom{i\gm^{\rm param,IR}_{\m_1\m_2\m_3\m_4}\times}
  {\ds +\, \big( \etaonethreetwofour-\etaonetwothreefour \big)
           \sin\!\left( \frac{p_1\wedge p_4}{2}\right) 
           \sin\!\left( \frac{p_3\wedge p_2}{2}\right) 
           \ln\vt }\\[12pt]
\hphantom{i\gm^{\rm param,IR}_{\m_1\m_2\m_3\m_4}\times}
  {\ds +\, \big( \etaonethreetwofour-\etaonetwothreefour \big)
           \sin\!\left( \frac{p_4\wedge p_2}{2}\right) 
           \sin\!\left( \frac{p_3\wedge p_1}{2}\right) 
           \ln\vu\,\bigg]~,}
\end{array}
\label{4pointparIR}
\end{equation}
$a^{\rm param}$ being as in eq. (\ref{coeffnon}). Note that, despite
the fact that the paramagnetic contribution to the four-point function
is made of integrals $J(p)$, with $p=p_1+p_2,~p_2+p_3,~p_1+p_3$, whose
non-planar parts become singular for $\tp=0$, $i\gm^{\rm
param,IR}_{\m_1\m_2\m_3\m_4}(p_1,p_2,p_3,p_4)$ is finite for any
configuration (\ref{4pIRregion}). This finiteness can be explained as
a consequence of gauge invariance following the same lines as for the
diamagnetic contribution: see paragraph below
eq. (\ref{4pointdiamIR}). Also note that in the momentum region
(\ref{4plog}) the value of $i\gm^{\rm param,IR}_{\m_1\m_2\m_3\m_4}
(p_1,p_2,p_3,p_4)$ can be obtained from the UV divergent contribution
$i\gm^{\rm param,UV}_{\m_1\m_2\m_3\m_4} (p_1,p_2,p_3,p_4)$ by using the
identifications (\ref{identifications}), in accordance with UV/IR
mixing.

It remains to study the mixed contribution to the four-point
function. There are quite a few diagrams producing contributions
quartic in the background field with both diamagnetic and paramagnetic
vertices. It is straightforward to see, however, that their planar
parts are all UV finite by power counting and that their non-planar
parts are free of singularities for momenta (\ref{4pIRregion}), so we
conclude that
\begin{equation}
   \gm^{\rm mixed,UV}_{\m_1\m_2\m_3\m_4}(p_1,p_2,p_3,p_4) = 
   \gm^{\rm mixed,IR}_{\m_1\m_2\m_3\m_4}(p_1,p_2,p_3,p_4) = 0~.
\label{4mixed}
\end{equation}

We are now ready to give the complete UV divergent contribution
$i\gm^{\rm UV}_{\m_1\m_2\m_3\m_4}(p_1,p_2,p_3,p_4)$ to the four-point
function. It is the sum of the UV divergent diamagnetic and
paramagnetic contributions and takes the form
\begin{equation}
\begin{array}{l}
  {\ds i\gm^{\rm UV}_{\m_1\m_2\m_3\m_4}(p_1,p_2,p_3,p_4) = 
       \frac{i}{\ee}\; \left( a^{\rm diam}+a^{\rm param} \right) }\\[9pt]
\hphantom{i\gm^{\rm UV}_{\m_1\m_2\m_3\m_4} }
  {\ds \times\, \bigg[\, 
       \big( \etaonethreetwofour-\etaonefourtwothree \big) 
       \sin\!\left( \frac{p_1\wedge p_2}{2}\right) 
       \sin\!\left( \frac{p_3\wedge p_4}{2}\right)}\\[12pt]
\hphantom{i\gm^{\rm UV}_{\m_1\m_2\m_3\m_4}\times}
  {\ds +\, \big( \etaonethreetwofour-\etaonetwothreefour \big)
           \sin\!\left( \frac{p_1\wedge p_4}{2}\right) 
           \sin\!\left( \frac{p_3\wedge p_2}{2}\right) }\\[12pt]
\hphantom{i\gm^{\rm UV}_{\m_1\m_2\m_3\m_4}\times}
  {\ds +\, \big( \etaonethreetwofour-\etaonetwothreefour \big)
           \sin\!\left( \frac{p_4\wedge p_2}{2}\right) 
           \sin\!\left( \frac{p_3\wedge p_1}{2}\right) \bigg] }
\end{array}
\label{4pointUV}
\end{equation}
It follows that the MS counterterm (\ref{counterterm}) subtracts the
UV divergences in the four-point function and that the beta function
computed from the resulting renormalized four-point is as in
eq. (\ref{nonbeta}).  As concerns the leading non-commutative IR
contribution to the four-point function, it is finite after summing
over diagrams and is given by the sum of eqs. (\ref{4pointdiamIR}) and
(\ref{4pointparIR}):
\begin{equation}
   i\gm^{\rm IR}_{\m_1\m_2\m_3\m_4} (p_1,p_2,p_3,p_4) 
   = i\gm^{\rm diam, IR}_{\m_1\m_2\m_3\m_4} (p_1,p_2,p_3,p_4) 
   + i\gm^{\rm param, IR}_{\m_1\m_2\m_3\m_4} (p_1,p_2,p_3,p_4)~.
\label{4pointIR}
\end{equation}

\section{Gauge invariance}
In this section we shall prove that gauge invariance
for the one-loop renormalized theory as mathematically expressed by
the equation
\begin{equation}
  \idxd\,\om(x)\, D_\m[B]\frac{\d\gm[B]}{\d B_\m (x)} = 0
\label{gaugeinvariance}
\end{equation}
holds, where $i\Gamma[B]=\ln Z[B]$ and $Z[B]$ is given in
eq. (\ref{uonepartition}).  In this regard it is worth noting that,
to the best of our knowledge, it remains an open question
\cite{Chepelev} whether non-logarithmic non-commutative IR divergences
are compatible with gauge invariance and finiteness. Here we see that
non-logarithmic IR divergences preserve gauge invariance.

We have shown in sections 2, 3 and 4 that, to subtract the UV
divergences in the two, three and four-point functions, it is enough
to add the counterterm (\ref{counterterm})
\begin{equation}
   \d S[B] = \frac{1}{4\ee} \left( a^{\rm diam}+a^{\rm param} \right) 
             \idxd\, F^{\m\n}[B] \star F_{\m\n}[B] ~.
\label{counterterm-5}
\end{equation} 
This counterterm satisfies eq. (\ref{gaugeinvariance}) and does not
change the non-commutative IR behaviour of the Green functions. Hence,
the renormalized two, three and four-point functions have the same
non-commutative IR behaviour as the regularized ones, given in
eqs. (\ref{vacuumIR}), (\ref{threepIR}) and (\ref{4pointIR}). We must
check that these expressions are consistent with the Ward identity
(\ref{gaugeinvariance}). To this end, we note that
eq. (\ref{gaugeinvariance}) implies 
\begin{eqnarray}
    & p^\n\,\Pi_{\m\n}(p) = 0 & \label{transverse2} \\[12pt]
    & {\ds p_3^{\m_3}\,\Gamma_{\m_1\m_2\m_3}(p_1,p_2,p_3) 
        = 2i\,\sin\!\left(\frac{p_1\wedge p_2}{2}\right) 
          \Big[ \Pi_{\m_1\m_2}(p_1)-\Pi_{\m_1\m_2}(p_2) \Big] ~,} &
                                 \label{transverse3} 
\end{eqnarray}
together with eq. (\ref{fourthreegaugeeq}). It is apparent that the
non-commutative IR divergent contribution $\Pi^{\rm IR}_{\m\n}(p)$ to
the vacuum polarization tensor in eq. (\ref{vacuumIR}) verifies the
identity (\ref{transverse2}). It is also clear that, neglecting terms
which remain finite for some $i$ $(i=1,2,3)$ in the limit $\tp_i\to
0$, we have
\begin{displaymath}
\begin{array}{l}
  {\ds p_3^{\m_3}\,\Gamma_{\m_1\m_2\m_3}^{\rm IR}(p_1,p_2,p_3) 
       \approx -\,\frac{2}{\pi^2}\,(p_1\wedge p_2)\, 
       \bigg[ \frac{(\tp_1)_{\m_1}(\tp_1)_{\m_2}}{\tp_1^2} 
            - \frac{(\tp_2)_{\m_1}(\tp_2)_{\m_2}}{\tp_2^2} \bigg] 
  }\\[12pt]
\hphantom{p_3^{\m_3}\,\Gamma_{\m_1\m_2\m_3}^{\rm IR}(p_1,p_2,p_3)~}
  {\ds \approx 2i \sin\!\left(\frac{p_1\wedge p_2}{2}\right) \,
       \Big[ \Pi_{\m_1\m_2}^{\rm IR}(p_1) 
           - \Pi_{\m_1\m_2}^{\rm IR}(p_2)\, \Big] ~,} 
\end{array}
\end{displaymath}
in agreement with the Ward identity (\ref{transverse3}) above. As
concerns the leading non-commutative IR behaviour of the four-point
function, its consistency with gauge invariance has already been
thoroughly checked in section 4.

We end this section writing the effective action which reproduces the
logarithmic contributions in the non-commutative IR momentum regions
(\ref{2plog}), (\ref{3plog}) and (\ref{4plog}) to the vacuum
polarization tensor, three-point function and four-point function.
{}From the results derived in sections 2, 3 and 4 it is clear that it
has the form
\begin{equation} 
   \frac{1}{4}\; \left(a^{\rm diam}+a^{\rm param}\right) 
   \,\ln(\theta\Lambda_{\rm IR})^2 \idx 
   \big(F^{\m\n}[B] \star F_{\m\n}[B] ~,
\label{IRren}
\end{equation} 
with the already known result (\ref{coeffnon}) for $a^{\rm diam}$ and
$a^{\rm param}$. In accordance with ref. \cite{MRS}, it is dual to the
UV counterterm (\ref{counterterm-5}) under the identifications
(\ref{identifications}).

\section{Conclusions and suggestions}

In this paper we have computed the complete UV and non-commutative IR
divergent contributions to the one-loop effective action of $U(1)$
gauge theory on non-commutative Minkowski space-time in a background
field. The UV divergences arise from the planar parts of the one-loop
diagrams, and come from virtual quanta with arbitrarily high momenta
moving around the loop of the diagrams.  The non-commutative IR
divergent contributions are also produced by high-momentum virtual
quanta, but occur in the non-planar parts of the diagrams. The vacuum
polarization has quadratic and logarithmic non-commutative IR
divergent contributions, while the three-point function has linear and
logarithmic contributions, and the four-point function only has
logarithmic contributions \cite{H} \cite{MST}. Logarithmic contributions
add to an overall IR finite contribution for the three and four-point
functions, while this is not so for the vacuum polarization tensor.
The overall logarithmic non-commutative IR behaviour of the three
Green functions is dual to the overall UV divergent behaviour in the
sense that the identifications (\ref{identifications}) transform one
into another.

The main result presented in this article is the following.  We have
shown that the paramagnetic contributions to the effective action
dominate over the diamagnetic contributions at very high
momentum. Here we say that we are in the high-momentum or high-energy
region if the momentum is large as measured against the
non-commutative energy scale $|\theta|^{-1/2}$. It turns out that, in
the high-momentum regime, paramagnetic contributions give rise to
anti-screening of the coupling constant, whereas diamagnetic
contributions produce screening. The combined effect explains the
negative sign of the beta function and its actual value. We recall
that, while for a Lorentz invariant theory paramagnetism always comes
with charge anti-screening and diamagnetism with charge screening
\cite{antiscreen2}, this is not necessarily so if the theory lacks
Lorentz invariance, a circumstance that occurs if space is
non-commutative. Yet the net result is analogous to that for
Yang-Mills theory on commutative Minkowski space-time. The reason for
this is that the beta function can be computed from the UV
divergences in the vacuum polarization tensor, provided gauge
invariance holds. These divergences occurring only in the non-planar
part implies that they are independent of $\theta^{\m\n}$ and do not
break Lorentz invariance.

We have also shown that the quadratic and linear non-commutative IR
divergences that exhibit the vacuum polarization tensor and
three-point function have a purely diamagnetic origin. By contrast,
the leading non-commutative IR logarithmic contributions to the vacuum
polarization tensor, the three-point and the four-point functions
receive both paramagnetic an diamagnetic contributions.  The duality
of logarithmic contributions to UV divergences discussed above does
not need to sum over diamagnetic and paramagnetic contributions but
holds for each of them separately.

Let us now look at the renormalized theory and consider the
non-commutative IR regime (\ref{4plog}). As shown in section 5, the
leading non-commutative IR logarithmic contribution to the effective
action in this region has the form (\ref{IRren}). This contribution
prevails over logarithmic contributions $\ln(p^2/\m^2)$, if $p^2$ is
of the same order of magnitude as the squared renormalization scale
$\m^2$.  This leads us to claim that, in this regime,
eq. (\ref{IRren}) can be thought of as giving rise to an effective IR
renormalization of the renormalized coupling constant of the theory
$g^{2\!}(\mu)$, thus introducing an IR effective coupling $g^2_{\rm
IR}$ given by
\begin{displaymath}
   \frac{1}{g^2_{\rm IR}} = \frac{1}{g^2(\mu)} 
       - \left( a^{\rm diam}+a^{\rm param}\right) \,
         \ln(\theta\Lambda_{\rm IR}\mu)^2,
\end{displaymath}
with $a^{\rm diam}$ and $a^{\rm param}$ as in eq. (\ref{coeffnon}).
In other words, in the non-commutative IR domain (\ref{4plog}), the
dominant logarithmic contribution to the 1PI one-loop Green functions
of the theory are given by their tree-level expressions at finite
$\theta$ but with $g^2(\m)$ replaced with $g_{\rm IR}^2$. Note also
that, in this non-commutative IR region, the paramagnetic
contributions produce a screening of the charge, whereas the
diamagnetic contributions anti-screen the charge. This does not come
as a surprise since we know that when Lorentz invariance is lost there
are media where paramagnetism comes with screening and diamagnetism
with anti-screening \cite{Ziman}, the dpeed of light being not equal
to 1. This is precisely what happens in the case at hand. As pointed
out in ref. \cite{MST}, the quadratic non-commutative IR divergent
behaviour of the vacuum polarization tensor gives rise to a
modification of the dispersion relation for photon polarizations
parallel to $\tilde{p}^{\mu}$.  Indeed, let $p^{\m}=(E,\vec{p})$, with
$\vec{p}=(p^1,p^2,p^3)$, and take a photon polarized along the
direction defined by $\tilde{p}^{\m}$. We then have
\begin{equation}
   E^2 = (p^3)^2 + \vec{P}^2\, \bigg( 1 
       - \frac{2}{\pi^2} \frac{g^2}{\theta^2|\vec{P}|^4}\bigg) ~,
\label{dispersion}
\end{equation}
where $\vec{P}=(p^1,p^2,0)$. This equation leads to a photon speed
$\,\vec{v}_g = \vec{\nabla}_{\!\!\vec{p}}\,E\,$, with modulus greater
than 1 in the region $2g^2\ll\pi^2\,\theta^2 |\vec{P}|^4$ where
perturbation theory is valid. Whether this signals a true instability
of the theory or is an artifact of perturbation theory remains an open
question \cite{Gomis} \cite{Landsteiner}. 

One may wonder if the results presented here for $U(1)$ gauge theory
generalize to the $U(N)$ case. In this regard we note that, although
in the non-commutative IR domain no new divergent dependence on the
external momenta appears, the colour structures of the UV and the
non-commutative IR divergent parts are different \cite{Armoni}
\cite{MRS}. This indicates that UV/IR duality in this case is more
involved and deserves further investigation. Finally, it will be worth
exploring whether there exists a connection between the results
presented here and the dipole structures that occur in non-commutative
quantum field theories \cite{dipole}.

\section*{Acknowledgement}

The authors are grateful to E. L\'opez for pointing them out that the
relative sign in the dispersion relation (\ref{dispersion}) in a
previous version of this paper was wrong. They also thank CICyT, Spain
for financial support through grant No. PB98-0842.

\newpage
\begin{center}
\epsfig{file=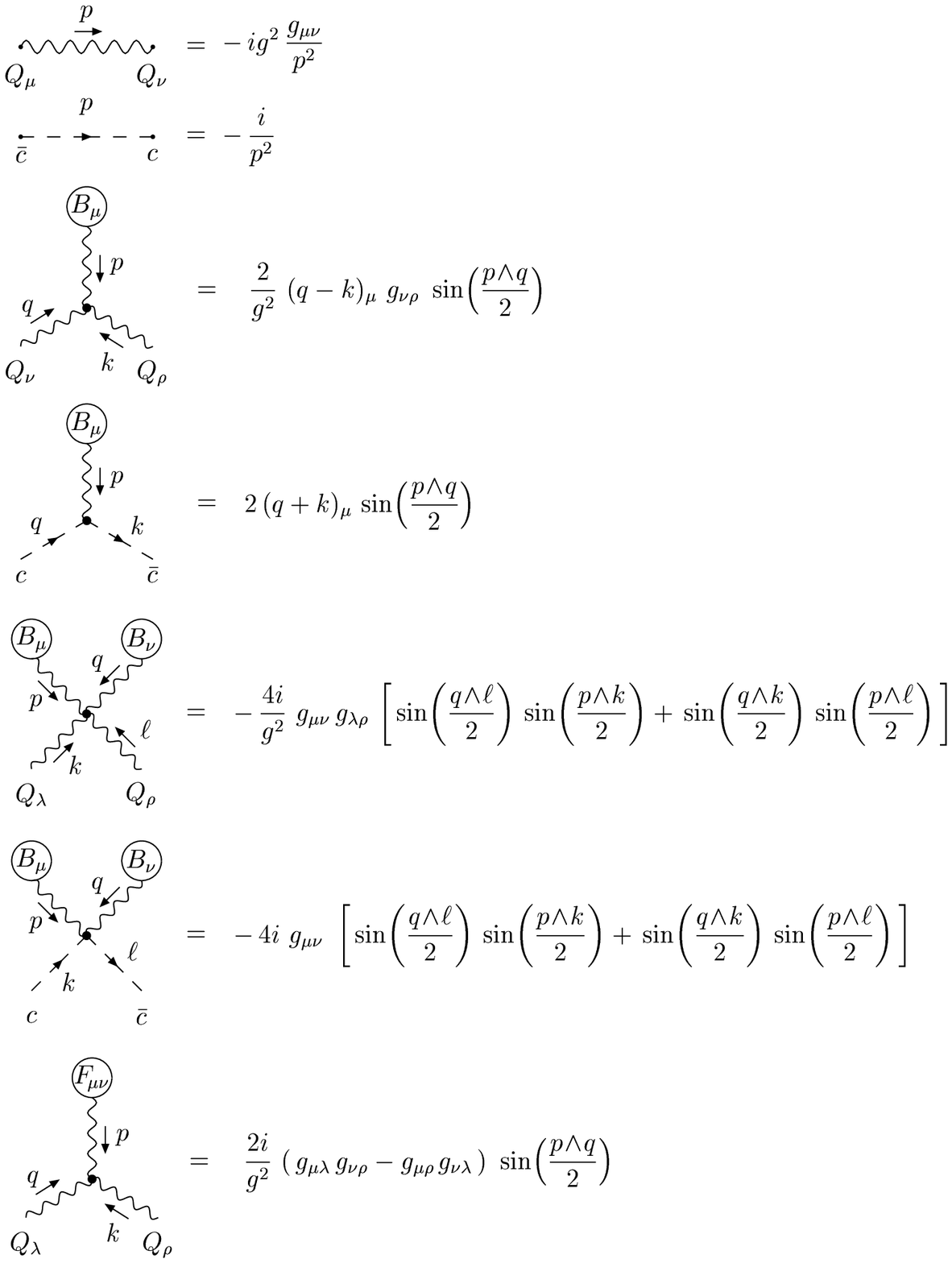,width=.85\textwidth} 
\end{center}
{\leftskip=40pt \rightskip=40pt 
{\sl Figure 1: Feynman rules of $U(1)$ gauge theory on
non-commutative Minkowski space-time in the Feynman background
gauge. The last vertex is the only one of paramagnetic type.}\par}

\newpage
\begin{center}
\epsfig{file=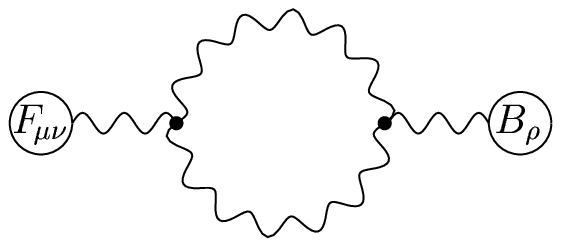,width=.30\textwidth} 
\end{center}
\vskip 15pt
{\leftskip=40pt \rightskip=40pt {\sl Figure 2: One-loop Feynman
diagram mixing diamagnetic and paramagnetic vertices contributing to
the vacuum polarization tensor.}\par}

\newpage
\begin{center}
\epsfig{file=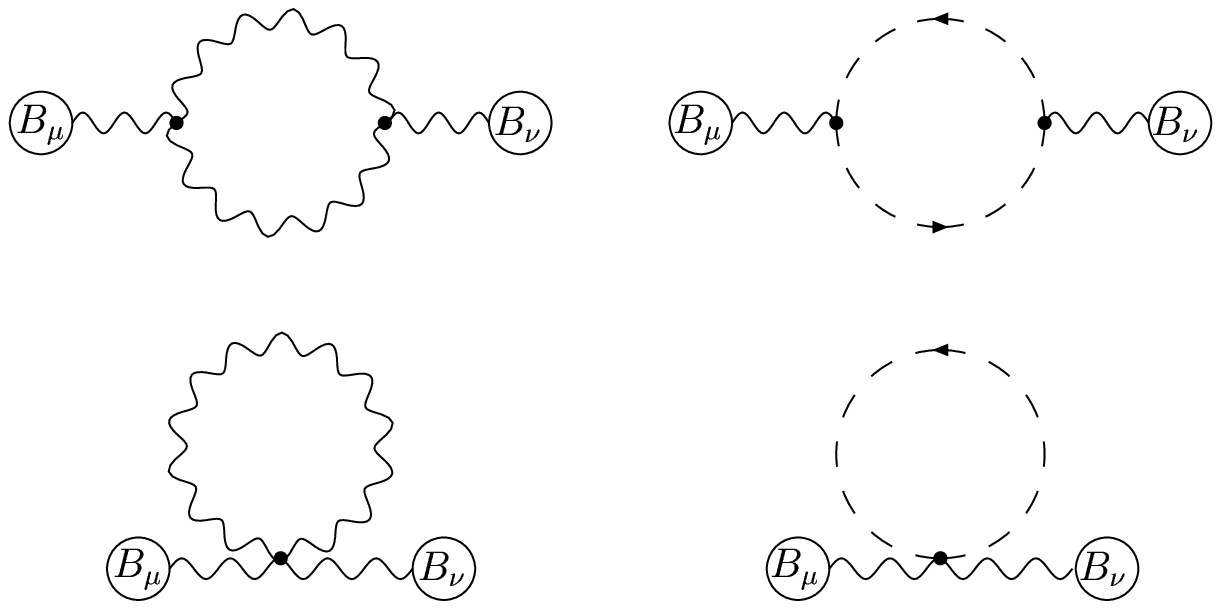,width=.70\textwidth} 
\end{center}
{\leftskip=40pt \rightskip=40pt 
{\sl Figure 3: One-loop diamagnetic Feynman diagrams contributing to
the vacuum polarization tensor. The diagrams include planar and
non-planar terms.}\par} 

\newpage
\begin{center}
\epsfig{file=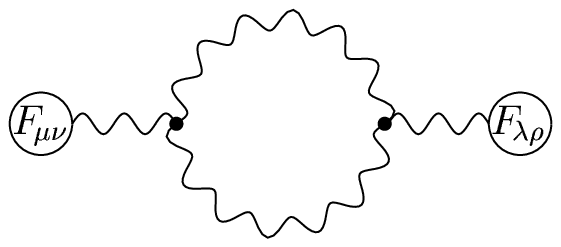,width=.30\textwidth} 
\end{center}
\vskip 15pt
{\leftskip=40pt \rightskip=40pt 
{\sl Figure 4: One-loop paramagnetic diagram contributing to the 1PI Green
function $\langle F_{\m\n}(-p) F_{\la\r}(p)\rangle$.
The diagram includes planar and non-planar terms.\par}}

\newpage
\begin{center}
\epsfig{file=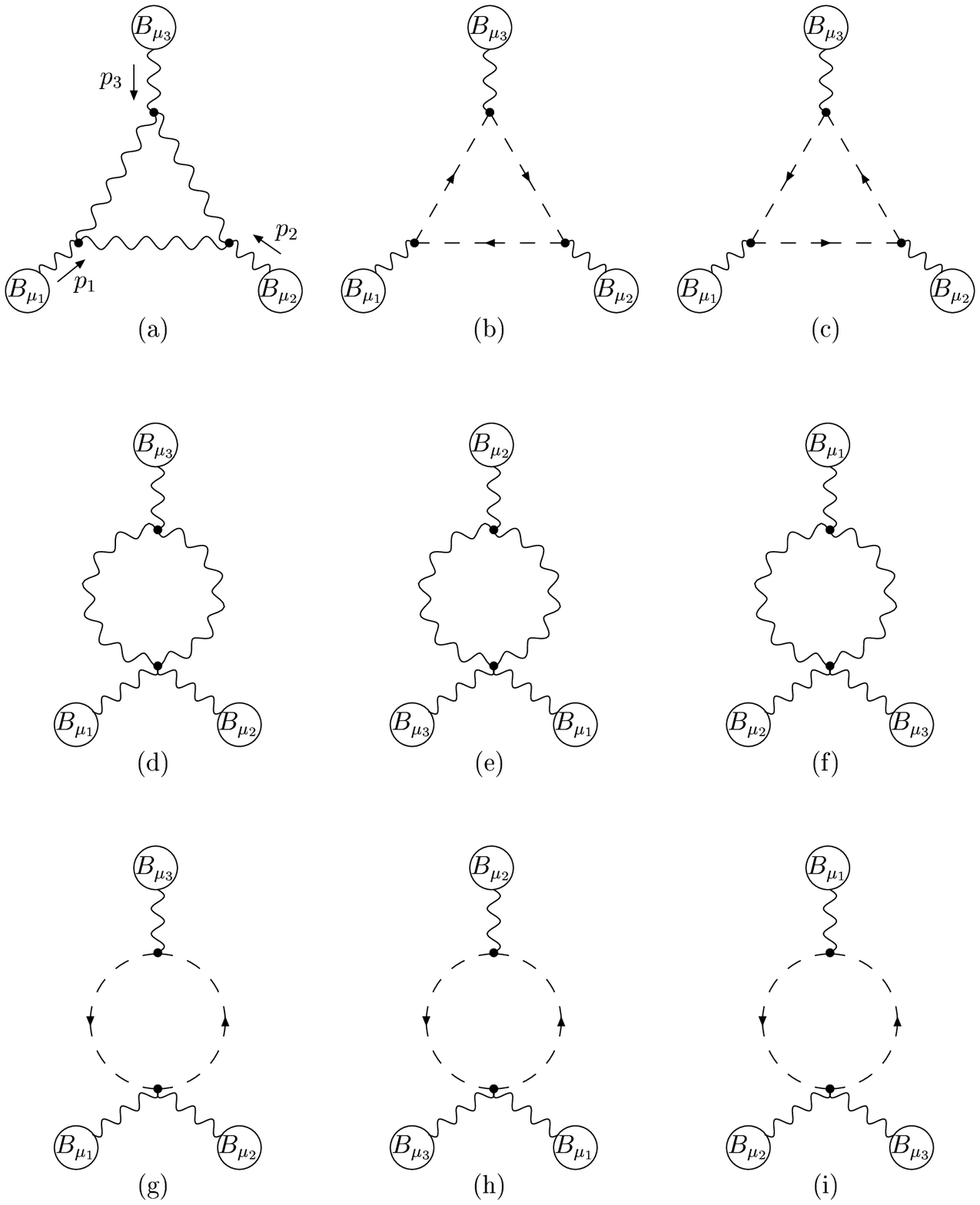,width=.85\textwidth} 
\end{center}
{\leftskip=40pt \rightskip=40pt 
{\sl Figure 5: One-loop 1PI diamagnetic diagrams contributing to the
three-point function.}

\newpage
\begin{center}
\epsfig{file=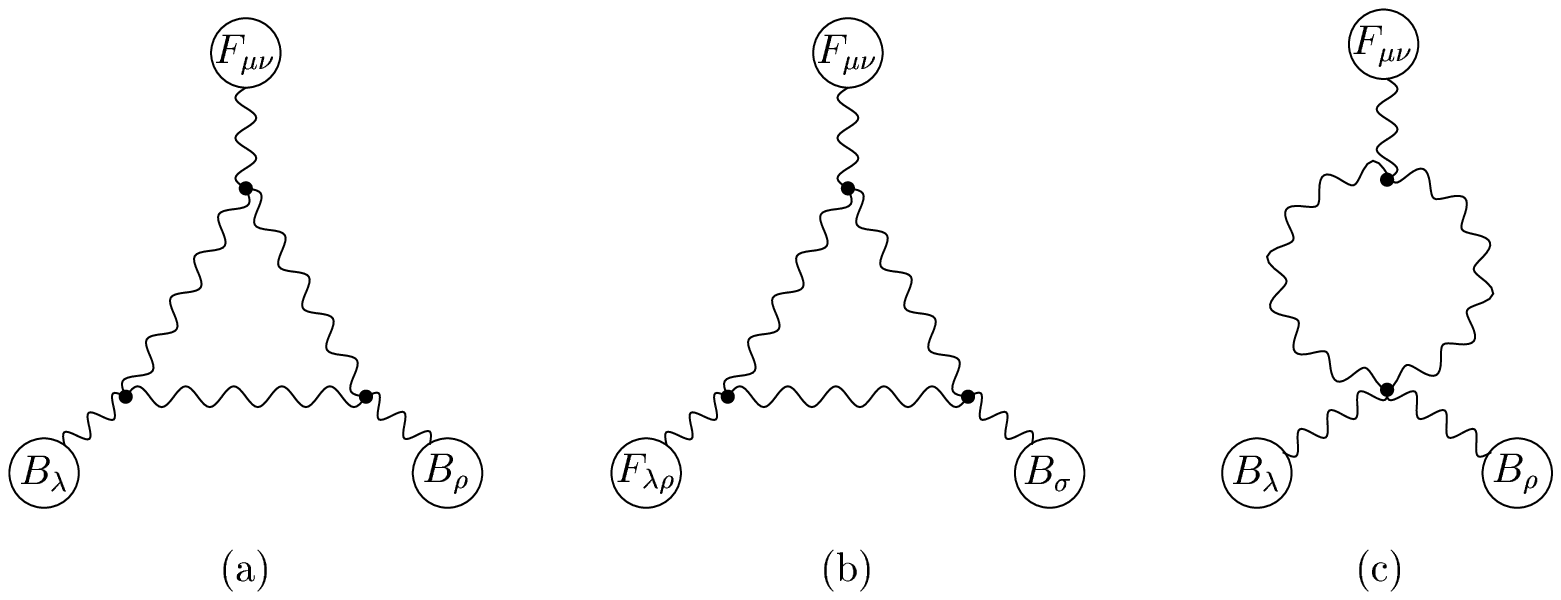,width=.85\textwidth} 
\end{center}
{\leftskip=40pt \rightskip=40pt 
{\sl Figure 6: One-loop mixed diagrams contributing to
the three-point function.}\par}

\newpage
\begin{center}
\epsfig{file=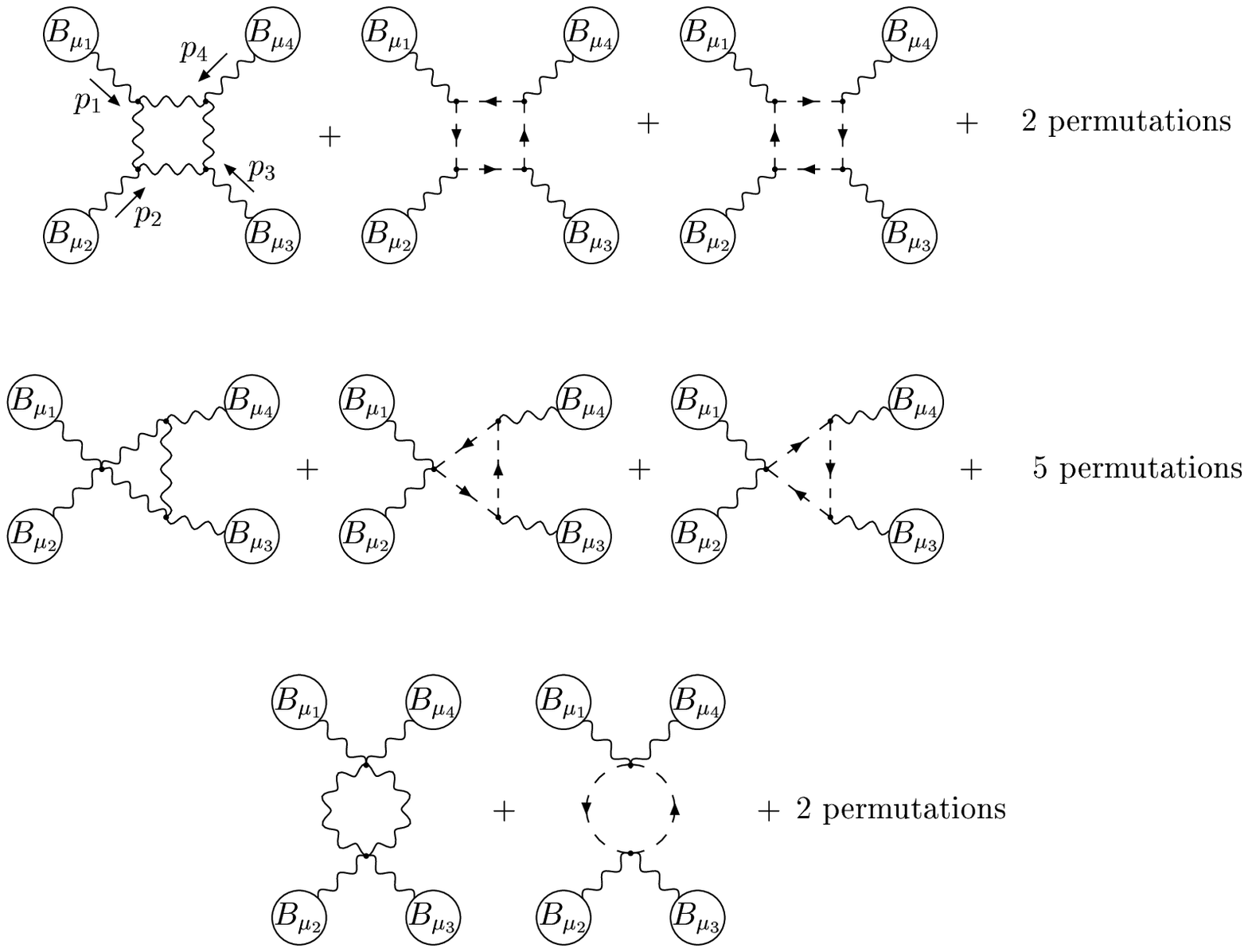} 
\end{center}
{\leftskip=40pt \rightskip=40pt 
{\sl Figure 7: One-loop 1PI diamagnetic diagrams contributing to the
four-point function.\par}}

\end{document}